\documentclass[twocolumn,showpacs,preprintnumbers,amsmath,amssymb]{revtex4-1} \usepackage{dcolumn}

\usepackage{graphicx} \usepackage{bm}

\newcommand{\bEw}{\bar{E}_w}
\newcommand{\bcE}{\bar{\cal E}}
\newcommand{\bcZ}{\bar{\cal Z}}

\newcommand{\cW}{{\cal W}}
\newcommand{\cZ}{{\cal Z}}

\newcommand{\hg}{\hat{g}}
\newcommand{\hga}{\hat{\gamma}}
\newcommand{\hG}{\hat{G}}
\newcommand{\hH}{\hat{H}}

\begin{document}


\title{Theory of Friedel oscillations in monolayer graphene and group-VI dichalcogenides in a magnetic field}
\date{\today}
\author{Tomasz M. Rusin}
\email{tmr@vp.pl}
\author{Wlodek Zawadzki}
\affiliation{Institute of Physics, Polish Academy of Sciences, Al. Lotnik\'ow 32/46, 02-688 Warsaw, Poland}

\begin{abstract}
Friedel oscillations~(FO) of electron density caused by a delta-like neutral
impurity in two-dimensional~(2D) systems in a magnetic field are calculated.
Three 2D cases are considered: free electron gas, monolayer graphene and
group-VI dichalcogenides. An exact form of the renormalized Green's function is used
in the calculations, as obtained by a summation of the infinite Dyson series
and regularization procedure.
Final results are valid for large ranges of potential strengths~$V_0$,
electron densities~$n_e$, magnetic fields~$B$ and distances from the impurity~$r$.
Realistic models for the impurities are used. The first FO of induced density in
WS$_2$ are described by the relation~$\Delta n({\bm r}) \propto \sin(2\pi r/T_{FO})/r^2$,
where~$T_{FO} \propto 1/\sqrt{E_F}$. For weak impurity potentials, the amplitudes of FO
are proportional to~$V_0$. For attractive potentials and high fields the total
electron density remains positive for all~$r$. On the other hand, for low fields,
repulsive potentials and small~$r$, the total electron density may become negative,
so that many-body effects should be taken into account.
\end{abstract}

\maketitle

\section{Introduction}

Disturbing a free-electron gas in a metal or
a semiconductor with an impurity gives rise to the Friedel oscillation~(FO)
of electron density~\cite{Friedel1952}. In the vicinity of the impurity,
usually a foreign atom embedded in the host material
or a vacancy, the electron density oscillates
\begin{equation} \label{Friedel_D}
 n(\bm r) \simeq n_0 + \delta n \frac{\sin(2k_Fr + \phi)}{r^D},
\end{equation}
where~$n_0$ is the electron gas density in the absence of
impurity,~$\delta n$ is the magnitude of induced density,~$k_F$
is the Fermi vector of the electron gas,~$D$ is dimensionality of the
system, and~$\phi$ is the phase shift.

Physically, FO result from a redistribution of
electrons caused by the potential of the impurity.
Because in a metal or a semiconductor only electrons with energies near
to the Fermi level can participate in the redistribution, the
induced density is characterized by the wave vector~$k\simeq k_F$.
Since their discovery in 1952, FO have been investigated
both theoretically and experimentally in many systems. For a
recent review of this subject see Ref.~\cite{Villain2016}.

Friedel oscillations were investigated in monolayer~\cite{Cheianov2006}
and bilayer graphene~\cite{Hwang2008} with the use of~${\bm k} \cdot {\bm p}$ and
tight binding methods~\cite{Bena2008,Bacsi2010}. Recently, they were
analyzed in hexagonal-lattices systems of
group-VI dichalogenides~\cite{Scholz2013}
and black phosphorous~\cite{Zou2016}. In all above papers the oscillations
were treated in absence of external fields. However, FO
also exist in the presence of a magnetic field, and their theoretical
investigation in 2D electron gases is a subject of the present work. Since
the problem of FO in a magnetic field is less frequently
discussed in the literature, we present a short review of this subject.

The first attempts to analyze FO in a magnetic
field were carried out by Rensink~\cite{Rensink1968},
Glasser~\cite{Glasser1969,Glasser1970} and Horing~\cite{Horing1969a,Horing1969b}
who considered 3D electron gas and delta-like or screened Coulomb
impurities. The main conclusions of these papers were:
i) for weak perturbations by an external potential, FO
follow formula~(\ref{Friedel_D}) with~$\phi=0$, ii) there is a qualitative
difference between oscillations in directions parallel and perpendicular to
magnetic field, iii) in strong quantizing fields the induced density
decays exponentially, iv) FO induced by the delta-like potential
are similar to those induced by the short-range screened Coulomb potential.

There exist several papers treating FO in a magnetic field in different systems.
Sedrakyan {\it et al.}~\cite{Sedrakyan2007} considered the impact of magnetic
field on a high-density electron gas and found a correction to the electrostatic
potential and an additional distance-dependent phase shift.
Sharma and Reddy~\cite{Sharma2011}
considered electronic screening at densities of relevance to neutron
star crusts and found that the screened potential between two static
charges exhibits long-range FO parallel to magnetic field that
can possibly create rod-like structures in the magnetar crusts.
Simion and Giuliani~\cite{Simion2005} treated FO in 2D electron gas
for electrons occupying the lowest Landau level.
Horing and Liu~\cite{Horing2009} treated FO
arising from the delta potential in monolayer graphene,
but they were faced with divergencies in the one-electron Green's function.
Bena~\cite{Bena2016} calculated FO in monolayer graphene in a magnetic field in the
lowest order of Born approximation. A problem of free electrons in 2D and 3D
interacting with point impurities in a magnetic field was analyzed by
Avishai {\it et al.}~\cite{Avishai1993} using an approach similar to that presented in our paper.
However, among properties calculated in Ref.~\cite{Avishai1993} the authors did not consider
the Friedel oscillations of electron density in modern materials.

Our work has three purposes. First, we want
to extend the results by Rensink, Glasser and
Horing~\cite{Rensink1968,Glasser1969,Glasser1970,Horing1969a,Horing1969b}
to 2D electron systems based on honeycomb lattices as, e.g., monolayer
graphene and group-VI dichalogenides MoS$_2$ or WS$_2$.
Second, being inspired by the approach of
Horing and Liu~\cite{Horing2009}, we apply a regularization method used in the quantum field
theory to handle the divergencies of one electron Green's function.
This method allows us to sum up exactly
the Born series for the Green's function of 2D electrons in a magnetic field in the
presence of a delta-like impurity for arbitrary strength of the potential.
Third, we to calculate FO going beyond the perturbation scheme by using the exact
one-electron Green's functions.
We hope that theoretical results for materials with 2D honeycomb lattice will
encourage experimental observation of FO in graphene and group-VI dichalogenides in a
magnetic field.

The crucial point of our approach is a possibility of exact summation of
the Born series, which requires isolation of the divergent part in the one-electron
Green's function~(GF) and a renormalization of the coupling constant. This approach was
successfully applied to high field magneto-resistance at low temperatures in 3D systems
by Gerhards and Hajdu~\cite{Gerhards1971} and we adopted their method to the 2D electron gas.
It turns out that, for realistic material parameters, energies and magnetic fields,
the renormalized potential of the impurity is close to the original one, which justifies
the regularization procedure. However, the proper regularization of the one-electron
GF, as described in our paper, requires knowledge of the one-electron GF
and its analytical behavior at the origin.
For this reason, our approach could not be used in the past when this feature
was not known. Below we give a short review of works related to this subject.

Some years ago Dodonov {\it et al.}~\cite{Dodonov1975} calculated stationary GF for
a free 2D electron in a homogeneous magnetic field and obtained analytical
results in terms of the Whittaker functions. Similar problems were recently
investigated for low-dimensional systems~\cite{Gusynin1995,Gorbar2002,Murguia2010} and
GF was obtained as infinite sums of Laguerre polynomials.
Horing and Liu~\cite{Horing2009} obtained a propagator as an infinite sum and,
alternatively, as the
second solution of Bessel wave equation. A closed form of the propagator in monolayer
graphene in terms of the confluent hypergeometric function was obtained by Pyatkovskiy and
Gusynin~\cite{Pyatkovskiy2011} and Gamayun {\it et al.}~\cite{Gamayun2011}.
The present authors~\cite{Rusin2011}
calculated the propagator in monolayer and bilayer graphene and obtained results
in terms of Whittaker functions. Ardenghi {\it et al.}~\cite{Ardenghi2015} calculated
GF of graphene taking into account corrections caused by the
coherent potential approximation, while Gutierrez-Rubio {\it et al.}~\cite{Rubio2016}
calculated numerically magnetic susceptibility of graphene and MoS$_2$
within the GF formalism. Recently, the present authors
published online the results for one electron GF in
a magnetic field for group-VI dichalogeinides~\cite{Rusin2016},
and Horing~\cite{Horing2017} considered the one-electron GF
in the same materials in position and momentum representations.

Our approach is based on the~${\bm k} \cdot {\bm p}$ theory of graphene
and group-VI dichalogeinides in the vicinity of~${\bm K}$ and~${\bm K}'$ points
of the Brillouin zone (BZ). We assume a non-interacting
electron gas at~$T=0$. The broadening parameters of GF
are taken from experimental values reported in the literature and
we use realistic values of impurity potentials.
For graphene we take a model potential of the nitrogen-like impurity which is
a frequent dopant in this material. Since we were unable to find
analogous model potential for group-VI dichalogeinides, we take for this case
potentials of nickel atom and a vacancy used in ab-initio calculations of high-T$_c$
superconductors.
Finally, we mention that FO in~WS$_2$ in a magnetic
field calculated in the parabolic-band approximation follow exact
results of Dodonov {\it et al.}~\cite{Dodonov1975}
for the GF of the 2D electron gas, while the calculations of FO
in monolayer graphene follow the results obtained by the present authors
in Ref.~\cite{Rusin2011}.

Our paper is organized as follows. In Section~II we introduce the theory of
FO in 2D electron systems including a detailed description of
the regularization procedure. Section~III contains our main results and
Section~IV their discussion. The paper is concluded by the Summary.
In Appendices we discuss auxiliary problems related to our main subject.

\section{Theory}

We describe the theory of density oscillations of a non-interacting
electron gas in 2D systems in the presence of a neutral delta-like impurity
in a constant magnetic field. These oscillations are similar to FO of a 3D electron gas
in the presence of an impurity in absence of
fields. The density of electron gas is calculated from GF of
the system consisting of free electrons in a magnetic field and a neutral impurity.
We analyze four models: 2D gas of noninteracting electrons,
electron gas in the conduction bands of monolayer graphene and
in group-VI dichalogenides (e.g. WS$_2$)
in the parabolic and non-parabolic approximations.

The calculations for the four systems are similar but for
2D electron gas GFs entering to the calculations are scalars
while for remaining systems GFs are block-diagonal~$4 \times 4$
or~$8 \times 8$ matrices. The presence of matrices
introduces complications in intermediate stages and final formulas,
but it does not change main conclusions about the physical
nature of FO in a magnetic field.
For this reason we present detailed calculations
for 2D electron gas, while for the other systems we
only quote the main steps of calculations and final results.

\subsection{2D electron gas in a magnetic field}

In the presence of a magnetic field the Hamiltonian for a 2D electron
gas is:~$\hH = (\hat{\bm p} + e{\bm A})^2/(2m_0)$,
where~$\hat{\bm p}$ is electron's momentum,~${\bm A}$ is vector potential,~$m_0$ is
electron mass and~$e$ is its charge~($e>0$).
In the Landau gauge~${\bm A}=(-By, 0)$. Let~$L=\sqrt{\hbar/eB}$ be the
magnetic radius and~$\xi =y /L-k_xL$.
Defining the standard raising and lowering operators for the
harmonic oscillator:~$\hat{a}ˆ=(\xi +\partial /\partial \xi)/\sqrt{2}$
and~$\hat{a}^+=(\xi -\partial /\partial \xi)/\sqrt{2}$ we have
\begin{equation} \label{2D_hH}
 \hH= \hbar \omega_c (\hat{a}^+\hat{a}+1/2),
\end{equation}
with~$\omega_c=eB/m_0$. The eigenstates of~$\hH$ are~$E_n=\hbar\omega(n+1/2)$,
where~$n=0,1,\ldots$ is the Landau level number, and the eigenstates of~$\hH$ are:
\begin{equation} \label{2D_Psi}
 \Psi_{nk_x}({\bm \rho}) = \frac{e^{ik_xx}}{\sqrt{2\pi}}\phi_n(\xi),
\end{equation}
where~${\bm \rho} = (x,y)$,
$\phi_n(\xi) = (1/\sqrt{L})C_n {\rm H}_{n}(\xi)e^{-\xi^2/2}$,
in which~${\rm H}_{n}(\xi)$ are the
Hermite polynomials, and~$C_n=1/\sqrt{2^n n!\sqrt{\pi}}$ are
the normalization coefficients. The electron spin is omitted.

The GF for the Hamiltonian~(\ref{2D_hH}) is~$\hg_{2D} = (E - \hH)^{-1}$.
In the position representation this operator reads
\begin{equation} \label{g2D_0}
\hg_{2D}({\bm \rho}_1,{\bm \rho}_2) = \sum_{n=0}^{\infty}\int_{-\infty}^{\infty}
 \frac{\phi_n(x,\xi_1)\phi_n^*(x',\xi_2)}{E-E_n} dk_x.
\end{equation}
Performing the integration over~$k_x$ (see Appendix~A) one obtains
\begin{eqnarray}
\label{g2D_1}
 \hg_{2D}({\bm \rho}_1,{\bm \rho}_2,\bcE) &=& -A(\omega_c)
 \sum_{n=0}^{\infty} \frac{L_n(r^2)}{n+1/2-\bcE}, \\
\label{g2D_A}
 A(\omega)&=& \frac{e^{-r^2/2 +i\chi}}{2\pi\hbar\omega L^2},
\end{eqnarray}
where~$\bcE=E/(\hbar\omega_c)$,~$r^2 = ({\bm \rho}_1-{\bm \rho}_2)^2/(2L^2)$,~$\chi=(x_1-x_2)(y_1+y_2)/2L^2$
is the gauge-dependent phase factor, and~$L_n(z)$ are the Laguerre polynomials.
The sum in Eq.~(\ref{g2D_1}) can be expressed in terms of the Whittaker functions~\cite{Dodonov1975}.
Introducing notation~$\hg_{{\bm 1},{\bm 2}} =\hg_{2D}({\bm \rho}_1,{\bm \rho}_2,\bcE)$ one has
\begin{equation} \label{g2D_Ex}
 \hg_{{\bm 1},{\bm 2}} = -\frac{e^{i\chi}}{2\pi\hbar\omega_c L^2|r| } \cW_{\bcE}(r^2),
\end{equation}
where we define:~$\cW_{\kappa}(z) = \Gamma(1/2-\kappa)W_{\kappa,0}(z)$, while~$\Gamma(z)$ and~$W_{\kappa,0}(z)$
are the Euler gamma and the Whittaker functions~\cite{GradshteinBook}, respectively.
Note the change of sign in Eq.~(\ref{g2D_Ex}) compared with
Refs.~\cite{Dodonov1975,Rusin2011}.

\subsection{Born series summation and regularization procedure}

Consider the Dyson equation for a point-like impurity
potential~$V(\bm \rho)=V_0\delta(\bm \rho- \bm \rho_0)$,
where~${\bm \rho_0}= (x_0,y_0)$ is the position of the impurity.
Note that~$V_0$ has the dimensionality of [energy]~$\times$ [area].
In the position representation there is
\begin{equation} \label{hG_2D_0}
 \langle {\bm \rho}_1| \hG|{\bm \rho}_2 \rangle = \langle {\bm \rho}_1| \hg |{\bm \rho}_2 \rangle +
 \int \langle {\bm \rho}_1 |\hg| {\bm \rho}_3\rangle V({\bm \rho}_3)
 \langle {\bm \rho}_3 |\hG |{\bm \rho}_2 \rangle d^2{\bm \rho}_3,
\end{equation}
in which~$\hg$ is GF of free electron gas in Eq.~(\ref{g2D_Ex}),
and~$\hG$ is GF of free electron gas in the presence of point-like impurity.
Performing in Eq.~(\ref{hG_2D_0}) the integration over~${\bm \rho}_3$ one has
\begin{equation} \label{hG_2D_1}
 \hG_{{\bm 1},{\bm 2}} = \hg_{{\bm 1},{\bm 2}} + V_0 \hg_{{\bm 1},{\bm 0}} \hG_{{\bm 0},{\bm 2}},
\end{equation}
where we used notation:~$\hg_{{\bm 1},{\bm 2}}=\langle {\bm \rho}_1| \hg|{\bm \rho}_2 \rangle$
and~$\hG_{{\bm 1},{\bm 2}}=\langle {\bm \rho}_1| \hG|{\bm \rho}_2 \rangle$, c.f. Eq.~(\ref{g2D_Ex}).
Following Ziman and others~\cite{ZimanBook,Koster1954,Wolff1961,Clogston1962},
the above equation can be solved analytically by
setting~$\bm \rho_1 \rightarrow {\bm \rho_0}$. After short algebra one finds
\begin{equation} \label{hG_2D_Ex}
 \hG_{{\bm 1},{\bm 2}} = \hg_{{\bm 1},{\bm 2}} +
 V_0 \hg_{{\bm 1},{\bm 0}} \frac{1}{1- V_0\hg_{\bm 0,\bm 0}} \hg_{{\bm 0},{\bm 2}}.
\end{equation}
The one-electron GFs in Eq.~(\ref{hG_2D_Ex}) have poles
for energies~$E=E_n=\hbar\omega_c(n+1/2)$,
and they diverge for~${\bm \rho}_1 \rightarrow {\bm \rho}_0$.
The standard way to overcome the first problem is
to treat the energy as a complex variable by adding or subtracting the
imaginary part~$\pm i\eta$.
Then the resulting GF has finite values at~$E=E_n$, and
the energy levels are smeared around~$E_n$ over a finite width~$\pm \eta/2$.
Physically,~$\eta$ is a phenomenological constant characterizing scattering
processes occurring in real samples.
In many cases, as e.g. for the free electron gas in absence of fields,
the above receipt allows one to overcome the second problem mentioned
above, i.e. the divergence of~$\hg_{\bm 0,\bm 0}$,
see Refs.~\cite{ZimanBook,Koster1954,Wolff1961,Clogston1962}.

However, for the GF in Eq.~(\ref{g2D_Ex}) the above procedure
is {\it not} sufficient, as illustrated below. Turning to Eq.~(\ref{g2D_1})
and using notation from Eq.~(\ref{g2D_Ex}) we find
\begin{equation} \label{g2D_r0}
 \hg_{\bm 0,\bm 0}= -A(\omega_c) \sum_{n=0}^{\infty} \frac{1}{n+1/2 -\bcE +i\eta},
\end{equation}
since for~${\bm \rho}_1 \rightarrow {\bm \rho}_2$ there is~${\bm r}={\bm 0}$
and there is~$L_n(0)=1$ for all~$n$.
The series in Eq.~(\ref{g2D_r0}) is harmonic and it diverges for all~$\bcE$ and~$\eta$.
As shown in Appendix~D, other methods of smearing GF
by using different bell-like forms also lead to divergences in the
real part of~$\hg_{\bm 0,\bm 0}$.
The above considerations suggest a different way to calculate~$\hg_{\bm 0,\bm 0}$.
Below we propose to apply the regularization technique, similar to that outlined
in Ref.~\cite{Gerhards1971}.

Consider the function~$\hg_{\bm \zeta,\bm 0}$ in Eq.~(\ref{g2D_Ex})
in the limit~$|\zeta|\rightarrow 0$.
Omitting terms tending to unity we have
\begin{equation} \label{hg_zeta}
 \hg_{\bm \zeta,\bm 0} = -\frac{1}{2\pi\hbar\omega_c L^2|\zeta| }\Gamma(1/2-\bcE)W_{\bcE,0}(\zeta^2).
\end{equation}
For small arguments there is (see formula 13.14.19 in Ref.~\cite{dlmf2017})
\begin{equation} \label{Whitt_k0}
 W_{\kappa,0}(z) = -\frac{\sqrt{z}}{\Gamma(a)} \left[\ln(z) + \psi(a) + 2\gamma \right] + O(z^{3/2}\ln(z)),
\end{equation} 	
where~$a=1/2-\kappa$,~$\psi(z)$ is the digamma function~\cite{GradshteinBook},
and~$\gamma\simeq 0.577$ is the Euler-Mascheroni constant.
Next, by using Eqs.~(\ref{hg_zeta}) and~(\ref{Whitt_k0}) we isolate the
divergent~$\hg_{\bm \zeta,\bm 0}^{div}$ and the regular~$\hg_{\bm \zeta,\bm 0}^{reg}$
parts of~$\hg_{\bm \zeta,\bm 0}$. There is
\begin{eqnarray}
 \label{g2D_dr}
 \hg_{\bm \zeta,\bm 0} &=& \hg_{\bm \zeta,\bm 0}^{div} + \hg_{\bm \zeta,\bm 0}^{reg}, \\
 \label{g2D_div}
 \hg_{\bm \zeta,\bm 0}^{div} &=& \frac{2\ln(\zeta)}{ 2\pi\hbar\omega_c L^2}, \\
 \label{g2D_reg}
 \hg^{reg} \equiv \hg_{\bm \zeta,\bm 0}^{reg} &=& \frac{\psi(1/2-\bcE) + 2\gamma}{2\pi\hbar\omega_c L^2}.
\end{eqnarray}
Note that~$\hg_{\bm \zeta,\bm 0}^{div}$ diverges logarithmical with~$\zeta$.
This allows us to treat~$\zeta$ as a cut-off parameter, see Discussion.
The regular part of~$\hg_{\bm \zeta,\bm 0}$ in Eq.~(\ref{g2D_reg}) does not depend on~$\zeta$, so
we may take:~$\hg_{\bm \zeta,\bm 0}^{reg} = \hg_{\bm 0,\bm 0}^{reg} \equiv \hg^{reg}$.
The function~$\hg^{reg}$ is finite except at the poles~$\bcE= n+1/2$. By taking the
energy as a complex variable we find that, in the vicinity of poles,
the density of states (DOS) obtained from~$\hg^{reg}$ is accurately
described by the Lorentz function, see Appendix~C.
Equations~(\ref{g2D_dr})--(\ref{g2D_reg}) agree with results
of Avishai {\it et al.}~\cite{Avishai1993}.

From Eqs.~(\ref{g2D_Ex}) and~(\ref{g2D_div})--(\ref{g2D_reg}) we obtain
\begin{equation} \label{hG_2D_R1}
 \hG_{\bm 1,\bm 2} = \hg_{\bm 1,\bm 2} +
 \hg_{\bm 1,\bm 0}\frac{V_0}{1 - V_0\hg_{\bm \zeta,\bm 0}^{div} -V_0\hg^{reg}} \hg_{\bm 0,\bm 2}.
\end{equation}
Now we redefine~$V_0\rightarrow V_r$ in such a way that~$\hG_{\bm 1,\bm 2}$
does not include~$\hg_{\bm \zeta,\bm 0}^{div}$ explicitly
\begin{equation} \label{hG_2D_RR}
 \hG_{\bm 1,\bm 2} = \hg_{\bm 1,\bm 2} +
 \hg_{\bm 1,\bm 0} \frac{V_r}{1 - V_r \hg^{reg}} \hg_{\bm 0,\bm 2}.
\end{equation}
By using Eqs.~(\ref{hG_2D_R1}) and~(\ref{hG_2D_RR}) we find
\begin{equation} \label{V_RR}
 V_r = \frac{V_0}{1-V_0 \hg_{\bm \zeta,\bm 0}^{div}}.
\end{equation}
The redefined potential~$V_r$ depends on the unknown
parameter~$\zeta \rightarrow 0$, see Eq.~(\ref{g2D_div}).
However, since the delta-like potential used in our model is an
idealized form of realistic potentials of impurities and vacancies
or atomic nuclei interacting with the electron gas, one can
identify~$\zeta$ with an effective range of these potentials.
For typical material parameters, see Discussion,
there is:~$V_r \simeq V_0$ for impurities, vacancies etc.,
and~$V_r \simeq cV_0$ for atomic nuclei potentials,
with~$c$ being on the order of unity.
Equations~(\ref{g2D_reg}),~(\ref{hG_2D_RR}), and~(\ref{V_RR}) are the final result of
the regularization procedure for 2D electron gas.

\subsection{Monolayer graphene}

For monolayer graphene in the vicinity of the~${\bm K}$ and~${\bm K}'$
points in BZ the Hamiltonian of electron in a magnetic field is~\cite{Gusynin2007}
\begin{equation} \label{MLG_hH0}
 \hH = u\tau \hat\sigma_x {\pi}_x + u\sigma_y\hat{\pi}_y,
\end{equation}
where~$u=1 \times 10^6$~m/s is the electron velocity,~$\sigma_x, \sigma_y$ are the
Pauli matrices in the standard notation and~$\tau=\pm 1$ for the~${\bm K}$ and~${\bm K}'$,
respectively. Introducing the raising and lowering operators, and
taking~$\tau=+1$ (i.e. the~${\bm K}$ point) one has
\begin{equation} \label{MLG_hH1}
\hH = \left(\begin{array}{cc} 0 & -\hbar \Omega \hat{a} \\
 -\hbar \Omega \hat{a}^+ & 0 \end{array}\right),
\end{equation}
where~$\Omega =\sqrt{2} u/L$. For both~${\bm K}$ and~${\bm K}'$ points the electron energies
are:~$E_{n\epsilon}=\epsilon \hbar{\Omega}\sqrt{n}$,~$n=0,1,2,\ldots$, and~$\epsilon=\pm 1$.

For monolayer graphene in a magnetic field GF
operator is a~$4 \times 4$ block-diagonal
matrix:~$\hg_{ML}=\left(\begin{array}{cc} \hg^{K} & 0\\ 0 &\hg^{K'} \end{array}\right)$,
in which~$\hg^{K} =\left(\begin{array}{cc} \hg^{uu} & \hg^{ul} \\ \hg^{lu} & \hg^{ll}\end{array}\right)$
     and~$\hg^{K'}=\left(\begin{array}{cc} \hg^{ll} & -\hg^{lu} \\-\hg^{ul} & \hg^{uu}\end{array}\right)$~\cite{Rusin2011}.
The indexes~$u$ and~$l$ correspond to the upper or the lower components of the eigen-vectors, respectively.
Then one has
\begin{eqnarray}
 \label{gMLG_uu}
 \hg^{uu}({\bm \rho},{\bm \rho}', \bcE) &=& -\frac{\bcE e^{i\chi}}{2\pi\hbar\Omega L^2|r|}\cW_{\kappa_{u}}(r^2),\\
 \label{gMLG_ll}
 \hg^{ll}({\bm \rho},{\bm \rho}', \bcE) &=& -\frac{\bcE e^{i\chi}}{2\pi\hbar\Omega L^2|r|}\cW_{\kappa_{l}}(r^2),\\
 \label{gMLG_ul}
 \hg^{ul}({\bm \rho},{\bm \rho}', \bcE) &=& \frac{m_{ul} }{r^2} \bcE(\hg^{uu} - \hg^{ll}),
\end{eqnarray}
where~$\kappa_{u}=\bcE^2-1/2$ and~$\kappa_{l}=\bcE^2+1/2$.
In Eq.~(\ref{gMLG_ul}) we defined~$m_{ul}= [(y-y')+i(x-x')]/(\sqrt{2}L)$.
For~$\hg^{lu}$ one obtains the expression analogous to
that for~$\hg^{ul}$ in~(\ref{gMLG_ul}), but with~$m_{ul}$
replaced by~$m_{lu} = [(y'-y)+i(x-x')]/(L\sqrt{2})$.

For the~${\bm K}$ point, the GF of electron gas in monolayer
graphene in the presence of a neutral impurity is
\begin{widetext} \begin{equation} \label{hG_MLG}
 \hG_{\bm 1,\bm 2} = \left(\begin{array}{cc} \hg^{uu} & \hg^{ul} \\ \hg^{lu} & \hg^{ll}\end{array}\right)_{\!\!\bm 1\bm 2}
 + V_0\left(\begin{array}{cc} \hg^{uu} & \hg^{ul} \\ \hg^{lu} & \hg^{ll}\end{array}\right)_{\!\!\bm 1\bm 0}
 \left(\hat{I} - V_0 \hg_{\bm 0,\bm 0} \right)^{-1}
 \left(\begin{array}{cc} \hg^{uu} & \hg^{ul} \\ \hg^{lu} & \hg^{ll}\end{array}\right)_{\!\!\bm 0\bm 2},
\end{equation} \end{widetext}
where the elements~$\hg^{ab}$ (with~$a,b \in \{u,l\}$) are given in Eqs.~(\ref{gMLG_uu})--(\ref{gMLG_ul}).

For one-electron GFs in Eqs.~(\ref{gMLG_uu})--(\ref{gMLG_ul})
there is again a problem with the divergences of~$\hg_{\bm 0,\bm 0}$, since these
functions contain the Whittaker functions diverging
for~${\bm \rho}_1 \rightarrow {\bm \rho}_0$. Similarly to the
case of 2D electron gas, this difficulty can be solved by applying the regularization
procedure. The calculations are analogous to those for 2D electron gas, and we obtain
\begin{widetext} \begin{equation} \label{hG_MLG_RR}
 \hG_{\bm 1,\bm 2} = \left(\begin{array}{cc} \hg^{uu} & \hg^{ul} \\ \hg^{lu} & \hg^{ll}\end{array}\right)_{\!\!\bm 1\bm 2}
 + V_r\left(\begin{array}{cc} \hg^{uu} & \hg^{ul} \\ \hg^{lu} & \hg^{ll}\end{array}\right)_{\!\!\bm 1\bm 0}
 \left(\hat{I} - V_r \hg^{reg} \right)^{-1}
 \left(\begin{array}{cc} \hg^{uu} & \hg^{ul} \\ \hg^{lu} & \hg^{ll}\end{array}\right)_{\!\!\bm 0\bm 2},
\end{equation} \end{widetext}
where we retained only the regular parts of~$\hg_{\bm 0,\bm 0}$.
From Eqs.~(\ref{hg_zeta})--(\ref{Whitt_k0}) and~(\ref{gMLG_uu})--(\ref{gMLG_ul})
we find that~$\hg_{\bm 0,\bm 0}$
is a diagonal operator:~$\hg_{\bm 0,\bm 0}=\left(\begin{array}{cc} \hg_{uu}^{div} + \hg_{uu}^{reg}& 0
 \\ 0 & \hg_{uu}^{div} + \hg_{uu}^{reg} \end{array}\right)$,
in which
\begin{eqnarray}
\label{gMLG_div}
 \hg_{uu}^{div} = \hg_{ll}^{div} &=&2\frac{\bcE\ln(\zeta)}{2\pi\hbar\Omega L^2}, \\
 \label{MLG_reg_uu}
 \hg^{reg}_{uu} &=& \frac{\bcE\left[\psi(-\bcE^2 + 1) + 2\gamma\right]}{2\pi\hbar\Omega L^2} ,\\
 \label{MLG_reg_ll}
 \hg^{reg}_{ll} &=& \frac{\bcE\ \left[\psi(-\bcE^2) + 2\gamma\right]}{2\pi\hbar\Omega L^2}.
 \end{eqnarray}
For~$\zeta \rightarrow 0$ the nondiagonal elements of~$\hg^{reg}$ and~$\hg^{div}$
vanish, see Appendix~A.
On equating Eqs.~(\ref{hG_MLG}) and~(\ref{hG_MLG_RR}) we obtain two equations
for the diagonal elements of~$\hg_{\bm 0,\bm 0}$. In both cases~$V_r$ is given in Eq.~(\ref{V_RR}).

\subsection{Group-VI dichalogenides: non-parabolic bands}

In the two-band~${\bm k} \cdot {\bm p}$ theory, the Hamiltonian for the
group-VI dichalcogenides materials in the vicinity of~${\bm K}$ and~${\bm K'}$ points
of BZ in the absence of fields is~\cite{Xiao2012}
\begin{equation} \label{VI_hH0}
 \hH = \frac{a_l t}{\hbar}(\tau\hat{\sigma}_x \hat{p}_x + \hat{\sigma}_y \hat{p}_y) +
 \frac{\Delta}{2}\hat{\sigma}_z
-\lambda \tau \frac{\hat{\sigma}_z-1}{2} \hat{s_z},
\end{equation}
where~$\hat{\bm p}$ is the electron momentum,~$\tau=\pm 1$ is the valley index,~$a$ is
the lattice constant,~$t$ is the effective hopping integral,~$\Delta$ is the
energy gap,~$\lambda$ is the spin splitting at the valence band edge caused by
the spin-orbit interaction, and~$\hat{s}_z$ is the Pauli matrix for electron spin.
The Hamiltonian~(\ref{VI_hH0}) describes an~$8 \times 8$ operator consisting of
four uncoupled~$2\times 2$ blocks on the diagonal, and each block is characterized
by different combinations of~$\tau$ and~$s_z$ numbers.
The material parameters entering~(\ref{VI_hH0}) are listed in~\cite{Xiao2012}
and are quoted in Table~1.

The band structure obtained from Eq.~(\ref{VI_hH0}) consists of two pairs of
energy bands having four combinations of~$\tau$ and~$s_z$ numbers.
Bottoms of the conduction bands are located at the same
energy~$E_c = \Delta/2$, while tops of the valence bands are located at
two energies~$E_v^{\tau s_z}= - \Delta/2 -\tau s_z\lambda$. Each pair of energy bands
has a different energy gap:~$E_g^{\tau s_z} = \Delta \pm \tau s_z\lambda$, and
different effective masses of electrons in the conduction bands.

It is convenient to rewrite the Hamiltonian~(\ref{VI_hH0}) in a more symmetric form
\begin{equation} \label{VI_hH1}
 \hH = \frac{a_l t}{\hbar}(\tau\hat{\sigma}_x \hat{p}_x + \hat{\sigma}_y \hat{p}_y) +
 E_w \hat{\sigma}_z + E_{s_z},
\end{equation}
in which~$E_w=\Delta/2 - E_{s_z}$ is half of the energy gap between a given pair of bands,
and~$E_{s_z}=s_z\tau\lambda/2$ shifts the zero of energy scale.

\begin{table}
\label{Table_1}
\caption{Parameters used in the Hamiltonian~(\ref{VI_hH0}) for
        four group-VI dichalcogenides, after~\cite{Xiao2012}.}
\begin{tabular}{|c|c|c|c|c|c|}
\hline
Material &~$a_l$~(\AA) &~$\Delta$~(eV) &~t$~(eV)$ &~$2\lambda$~(eV) &~$a_l t/\hbar$ (10$^6$~m/s) \\
 \hline
  MoS$_2$ & 3.193 & 1.66 & 1.10 & 0.15 & 0.53 \\
  WS$_2$  & 3.197 & 1.79 & 1.37 & 0.43 & 0.66 \\
  MoSe$_2$& 3.313 & 1.47 & 0.94 & 0.18 & 0.47 \\
  WSe$_2$ & 3.310 & 1.60 & 1.19 & 0.46 & 0.60 \\
 \hline
\end{tabular}
\end{table}

In the presence of a magnetic field we replace in Eq.~(\ref{VI_hH1})
the electron momentum~$\hat{\bm p}$ by~$\hat{\bm \pi} = \hat{\bm p} +|e|{\bm A}$.
Taking the Landau gauge and introducing the raising and lowering operators,
as described for 2D case, we obtain for electrons at the~${\bm K}$ point
\begin{equation} \label{VI_hH1B}
 \hH = \left(\begin{array}{cc} E_w & -\hbar \Omega \hat{a} \\
 -\hbar \Omega \hat{a}^+ & -E_w \end{array}\right) + E_{s_z},
\end{equation}
where~$\Omega =\sqrt{2} a_lt/(\hbar L)$,~$E_{s_z}=s_z\tau\lambda/2$,~$s_z=\pm 1$, and~$\tau=+1$.
For~$n \geq 1$ the eigenenergies of the above Hamiltonian are
\begin{equation} \label{VI_E1}
 E_{nk_x \epsilon s_z} = \epsilon \sqrt{n \hbar^2 \Omega^2 + E_w^2} + E_{s_z}
 =\epsilon E_n + E_{s_z}.
\end{equation}
For~$n=0$ the eigenenergy is~$E_{0k_x,-1,s_z} = -E_w + E_{s_z}$, and there is no state
with~$n=0$ and~$\epsilon=+1$. Thus, at the~${\bm K}$ point, in the conduction bands
the lowest LL is~$n=1$, while in the valence bands the highest LL is~$n=0$.

At the~${\bm K}'$ point, the Hamiltonian of electron in a magnetic field is
\begin{equation} \label{VI_hH2B}
 \hH' = \left(\begin{array}{cc} E_w' & +\hbar \Omega \hat{a}^+ \\
 +\hbar \Omega \hat{a} & -E_w' \end{array}\right) + E_{s_z}',
\end{equation}
where~$E_{s_z}'=s_z\tau\lambda/2$ with~$\tau=-1$, and~$E_w'=\Delta/2 - E_{s_z}'$
is half the energy gap.
Note that~$\hH'= \sigma_z\hH^{T}\sigma_z$, see Eq.~(\ref{VI_hH1B}).
For~$n \geq 1$ the eigenenergies of the Hamiltonian~(\ref{VI_hH2B}) are
\begin{equation} \label{VI_E2}
 E_{nk_x \epsilon s_z}' = \epsilon \sqrt{n \hbar^2 \Omega^2 + E_w'^2} + E_{s_z}'
 =\epsilon E_n' + E_{s_z}'.
\end{equation}
For~$n=0$ the eigenenergy is~$E_{0k_x,+1,s_z}' = E_w' + E_{s_z}'$, and there is no state
with~$n=0$ and~$\epsilon=-1$. Thus, at the~${\bm K}'$ point, in the conduction bands
the lowest LL is~$n=0$, while in the valence bands the highest LL is~$n=1$. The above
asymmetry between the energies of Landau levels~$n=0$ at the~${\bm K}$ and~${\bm K}'$
points is discussed in detail in Ref.~\cite{Rose2013}.

For group-VI dichalcogenides,~$\tau=+1$ and for both orientations~$s_z$
the GF is~$\hg_+ =\left(\begin{array}{cc} \hg^{uu}_+ & \hg^{ul}_+ \\ \hg^{lu}_+ & \hg^{ll}_+ \end{array}\right)$,
where
\begin{eqnarray}
 \label{gVI_uu_p}
 \hg^{uu}_+({\bm \rho},{\bm \rho}',\cZ) &=& -\frac{(\bcZ +\bEw) e^{i\chi}}{2\pi\hbar\Omega L^2|r|} \cW_{\kappa_{u}}(r^2),\\
 \label{gVI_ll_p}
 \hg^{ll}_+({\bm \rho},{\bm \rho}',\cZ)
 &=& -\frac{(\bcZ -\bEw) e^{i\chi}}{2\pi\hbar\Omega L^2|r|} \cW_{\kappa_{l}}(r^2),\\
 \label{gVI_ul_p}
 \hg^{ul}_+({\bm \rho},{\bm \rho}',\cZ) &=&
 \frac{m_{ul}}{r^2}\left[(\bcZ-\bEw)\hg^{uu}_+ - (\bcZ+\bEw)\hg^{ll}_+ \right]\!, \ \ \
\end{eqnarray}
where~$\bcZ=(\cZ- E_{s_z})/(\hbar\Omega)$, and~$\bEw=E_w/(\hbar\Omega)$.
Here~$\kappa_{u}=\bcZ^2-\bEw^2 -1/2$ and~$\kappa_{l}=\bcZ^2-\bEw^2 +1/2$.
Details of derivation of Eqs.~(\ref{gVI_uu_p})--(\ref{gVI_ul_p}) are shown in Appendix~A.
For~$\tau=-1$ and for both orientations~$s_z$ there
is~$\hg_- =\left(\begin{array}{cc} \hg^{ll}_- & \hg^{lu}_- \\ \hg^{ul}_- & \hg^{ll}_- \end{array}\right)$,
where
\begin{eqnarray}
 \label{gVI_uu_m}
 \hg^{uu}_-({\bm \rho},{\bm \rho}',\cZ) &=& -\frac{(\bcZ -\bEw') e^{i\chi}}{2\pi\hbar\Omega L^2|r|} \cW_{\kappa_{u}}(r^2),\\
 \label{gVI_ll_m}
 \hg^{ll}_-({\bm \rho},{\bm \rho}',\cZ)
 &=& -\frac{(\bcZ +\bEw') e^{i\chi}}{2\pi\hbar\Omega L^2|r|} \cW_{\kappa_{l}}(r^2),\\
 \label{gVI_ul_m}
 \hg^{ul}_-({\bm \rho},{\bm \rho}',\cZ) &=&
 \frac{m_{ul}}{r^2}\left[(\bcZ - \bEw')\hg^{uu}_- - (\bcZ + \bEw')\hg^{ll}_- \right]\!, \ \ \
\end{eqnarray}
where~$\bcZ=(\cZ'- E_{s_z}')/(\hbar\Omega)$, and~$\bEw'=E_w'/(\hbar\Omega)$.
The GF of electron gas in group-VI dichalcogenides in the presence of a neutral impurity
is an~$8 \times 8$ block-diagonal matrix,
whose elements are combinations of~$\hg^{uu}_{\pm}$,~$\hg^{ll}_{\pm}$ and~$\hg^{lu}_{\pm}$ functions,
as given in Eqs.~(\ref{gVI_uu_p})--(\ref{gVI_ul_m}).
This matrix consists of four~$2\times 2$ blocks describing contributions
from~$\tau=\pm 1$ valleys and~$s_z=\pm 1$ spin orientations.

The regularization procedure for group-VI dichalogenides leads to equations analogous
to Eqs.~(\ref{hG_MLG})--(\ref{hG_MLG_RR}) for monolayer graphene, in which the
elements~$\hg^{ab}$ for~$a,b \in\{u,l \}$ are given in
Eqs.~(\ref{gVI_uu_p})--(\ref{gVI_ul_p}),~$V_r$ is defined in Eq.~(\ref{V_RR}),
and for~$\tau=+1$ there is
\begin{eqnarray}
 \label{gVI_div_uu}
 \hg^{div}_{uu+} &=& 2\frac{(\bcZ+\bEw)\ln(\zeta)}{2\pi\hbar\Omega L^2},\\
 \label{gVI_div_ll}
 \hg^{div}_{ll+} &=& 2\frac{(\bcZ-\bEw)\ln(\zeta)}{2\pi\hbar\Omega L^2},\\
 \label{gVI_reg_uu}
 \hg^{reg}_{uu+} &=& \frac{(\bcZ+\bEw)\left[\psi(\bEw^2-\bcZ^2 + 1) + 2\gamma\right]}{2\pi\hbar\Omega L^2} ,\\
 \label{gVI_reg_ll}
 \hg^{reg}_{ll+} &=& \frac{(\bcZ-\bEw)\left[\psi(\bEw^2-\bcZ^2) + 2\gamma\right]}{2\pi\hbar\Omega L^2},
\end{eqnarray}
while~$\hg^{reg}_{ul+}=\hg^{reg}_{lu+}=0$, see Appendix~A. For~$\tau=-1$ we obtain expressions
analogous given above, see Eqs.~(\ref{gVI_uu_m})--(\ref{gVI_ul_m}).

\subsection{Group-VI dichalogenides: parabolic bands approximation}

Consider the difference~$\delta E_n$ of energies~$E_{n+1,k_x \epsilon s_z}$
and~$E_{n,k_x \epsilon s_z}$ for the Hamiltonian in Eq.~(\ref{VI_hH1B}).
For energies in Eq.~(\ref{VI_E1}),~$\epsilon=+1$, and
magnetic fields~$B< 40$~T, there is:~$(\hbar \Omega)^2 \ll E_w^2$,
and one obtains
\begin{equation} \label{VI_dE1}
 \delta E_n = E_{n+1,k_x \epsilon s_z} - E_{n,k_x \epsilon s_z} =
 \frac{\hbar^2 \Omega^2}{E_{n+1}+E_n} \simeq \frac{\hbar eB u^2}{E_w},
\end{equation}
where~$n \geq 1$,~$u=a_l t/\hbar \simeq 0.66 \times 10^6$~m/s, see Table~1,
and~$E_w$ is defined in Eq.~(\ref{VI_hH1}).
The quantity~$1/m^*=u^2/E_w$ is the inverse of the energy effective mass
of electron at the bottoms of conduction bands.
Since~$E_w= \Delta/2 +\tau s_z \lambda$, see Eq.~(\ref{VI_hH1}), one obtains
two different values of effective masses, depending on the sign of~$\tau s_z$ product.
When these masses are large enough, we may approximate the system described by
the Hamiltonian~(\ref{VI_hH1B}) by a system of two
Hamiltonians of 2D free electron gases having
the cyclotron frequencies:~$\omega_c^+=eB/(\Delta + \lambda)$
for~$\tau s_z=1$ and~$\omega_c^- = eB/(\Delta - \lambda)$ for~$\tau s_z=-1$.
For example, for WS$_2$ at~$B=9$~T one has:~$\hbar\omega_c^+ = 2.62$~meV
and~$\hbar\omega_c^- = 3.33$~meV. In absence of the spin splitting~($\lambda=0$)
there is:~$\hbar\omega_c^{\pm}=\hbar\omega_c= 2.93$~meV.
We call this model a parabolic approximation for energy bands.

In the nonparabolic model of energy bands, at~${\bm K}'$ point
the Landau level with~$n=0$ is
at the bottom of the conduction bands, see Eq.~(\ref{VI_E1}).
To keep the same position of the Landau level with~$n=0$ in the parabolic case
one has to shift the zero of energy scale
from~$E=\Delta/2$ to~$E=\Delta/2 - \hbar\omega_c^{\pm}/2$.
At the~${\bm K}$ point the lowest LL in the conduction bands is~$n=1$. To be consistent
with the previous case we shift the zero of energy scale at~${\bm K}$ point to the same
value~$E=\Delta/2 - \hbar\omega_c^{\pm}/2$, but exclude LL with~$n=0$ from
further considerations.
Then, the energy spectrum of conduction electrons consists of
four ladders of Landau levels having two different cyclotron
energies~$\hbar\omega_c^{\pm}$, respectively. At the~${\bm K}'$ point ($\tau=-1$) the
ladders start from~$n=0$, while at the~${\bm K}$ point ($\tau=+1$) they start from~$n=1$.
In both cases the eigenstates are described by functions~$\Psi_{nk_x}({\bm \rho})$
in Eq.~(\ref{2D_Psi}).

To estimate the accuracy of Eq.~(\ref{VI_dE1}), we calculate~$\delta E_n$ for~WS$_2$
for~$B=9$~T taking~$\lambda=0$. For~$n=1$ there is a negligible difference
between~$\delta E_1$ and~$\hbar\omega_c = 2.93$~meV,
while for~$n=16$ the difference~$\delta E_{16}$ is only five percent smaller
than~$\hbar\omega_c = 2.93$. Since we concentrate on small~$n$ values,
one may safely use the parabolic approximation introduced above.

Within the parabolic approximation of energy bands, the one-electron GF
is analogous to that given in Eq.~(\ref{g2D_Ex})
with~$\omega_c \rightarrow \omega_c^{\pm}$,
and~$\bcE\rightarrow \bcZ = (E - \Delta/2)/\hbar\omega_c^{\pm} + 1/2$.
Thus we obtain
\begin{equation} \label{g2D_PA}
 \hg_{PA}^{\pm}({\bm \rho},{\bm \rho}', E) =
 -\frac{e^{i\chi}}{2\pi\hbar\omega_c^{\pm} L^2}\left( \frac{\cW_{\bcZ}(r^2)}{|r|}-
 \delta_{\tau,1}\frac{e^{-r^2/2}}{\bcZ} \right).
\end{equation}
The last term in Eq.~(\ref{g2D_PA}) arises from the exclusion of term~$n=0$ from the
summation, see Eq.~(\ref{g2D_1}).
In the parabolic model the regularization procedure leads to, see Eq.~(\ref{hG_2D_RR}),
\begin{equation} \label{hG_PA_RR}
 \hG_{\bm 1,\bm 2} = \hg_{\bm 1,\bm 2} +
 \hg_{\bm 1,\bm 0} \frac{V_r}{1 - V_r \hg^{reg}} \hg_{\bm 0,\bm 2},
\end{equation}
where~$\hg^{div}$,~$V_r$ and~$\hg$ are
defined in Eqs.~(\ref{g2D_div}),~(\ref{V_RR}),~(\ref{g2D_PA}), respectively, and
\begin{equation}
 \label{gPA_reg}
 \hg^{reg} = \frac{\psi(1/2-\bcZ) + 2\gamma- \delta_{\tau,1}/\bcZ }{2\pi\hbar\omega_c^{\pm} L^2}.
\end{equation}

To find the electron density from GF
in Eq.~(\ref{hG_PA_RR}) one has to sum contributions from two bands
having frequencies~$\omega_c^+$ and the other two bands having frequencies~$\omega_c^-$.
Performing this summation one should remember that the four energy bands are
filled up to {\it the same Fermi level}. This means that, at~$T=0$, each pair of
bands has a different index~$n_{max}$ of the top-most occupied Landau level.

Equations~(\ref{hG_2D_RR}),~(\ref{hG_MLG_RR}), and~(\ref{hG_PA_RR})
with~$V_r$ defined in Eq.~(\ref{V_RR}), as well as the one-electron Green's
functions~$\hg$ and their regularized parts~$\hg^{reg}$
are the final formulas for the Green's functions
of electrons in 2D systems in the presence of a magnetic field
and a delta-like impurity at the origin. These results are {\it exact}
since the corresponding Born series are summed up to all terms.
They may be used within the whole range of energies, parameters~$V_0$ and
distances from the impurity within the range of validity of
the one-electron approximation, see Discussion.

\subsection{Electron density}

In this section we calculate the electron density obtained
from~$\hG({\bm \rho}_1, {\bm \rho}_2)$ given in
Eqs.~(\ref{hG_2D_RR}),~(\ref{hG_MLG_RR}) and~(\ref{hG_PA_RR}).
For~$T=0$, the density of the electron gas~$n(\bm \rho)$ is
\begin{equation} \label{n_rho}
 n(\bm \rho)= -\frac{2}{\pi}\int_{E_x}^{E_F} {\rm Im} {\rm Tr}\{ \hG(\bm \rho,\bm \rho) \} dE,
\end{equation}
where~$E_F$ is the Fermi energy and~$E_x$ is a suitable cut-off energy in the valence band,
see Discussion. The part of charge density induced by the presence of impurity is
\begin{equation} \label{Delta_n}
 \Delta n(\bm \rho)= -\frac{2}{\pi}\int_{E_x}^{E_F}
 {\rm Im} {\rm Tr}\{ \hG(\bm \rho,\bm \rho) - \hg(\bm \rho,\bm \rho) \} dE,
\end{equation}
and this quantity is analyzed in detail below.
To estimate the magnitude of~$\Delta n(\bm \rho)$ in Eq.~(\ref{Delta_n}) we
linearize the expressions for~${\rm Tr}\{\hG(\bm \rho,\bm \rho)\}$
and obtain:~$\Delta n(\bm \rho) \sim \hg V_r \hg$.
For monolayer graphene we take:~$B=10$~T,~$V_r \simeq V_0 = -56$~eV$\AA^2$, see Discussion.
Then one obtains
\begin{eqnarray} \label{Delta_n_MLG}
 \Delta n(\bm \rho)&\propto &
 \left(\frac{V_r}{\hbar \Omega L^2} \right)
 \left(\frac{\rm [meV]}{\hbar\Omega L^2}\right) \times \Upsilon(\rho) \nonumber \\
 &=& 2.1\times 10^{9} {\rm cm^{-2}} \times \Upsilon^{ML}(\rho),
\end{eqnarray}
where~$\Upsilon^{ML}(\rho)$ is a dimensionless function, see Eq.~(\ref{Delta_n}).
For WS$_2$ we take:~$V_r \simeq V_0 =-4.3$~eV$\AA^2$ and obtain
\begin{eqnarray} \label{Delta_n_WS2}
 \Delta n(\bm \rho)&\propto & 9.4\times 10^{10} {\rm cm^{-2}} \times \Upsilon^{WS}(\rho),
\end{eqnarray}
where~$\Upsilon^{WS}(\rho)$ is an another dimensionless function.

For a given electron concentration~$n_e^{\pm}$ and a magnetic field~$B$,
as well as specified form of the density of states (DOS),
the Fermi level is calculated in an
unique way~\cite{Zawadzki1984}, see below. In our calculations we assume the
Lorentz-like DOS, as obtained from the regularized Green's functions~$\hg^{reg}$ for
the four systems under consideration, see Eq.~(\ref{g2D_reg_a}) in Appendix~C.

The local density of states (LDOS) can be obtained from GF as
\begin{equation} \label{LDOS}
 {\rm LDOS} = -\frac{2}{\pi} {\rm Im} {\rm Tr}\{ \hG(\bm \rho,\bm \rho) \}.
\end{equation}
This quantity was employed recently to determine FO in
several systems with the use of scanning tunneling microscopy, see Discussion.

\section{Friedel oscillations}

\begin{figure}
\includegraphics[width=8cm,height=8cm]{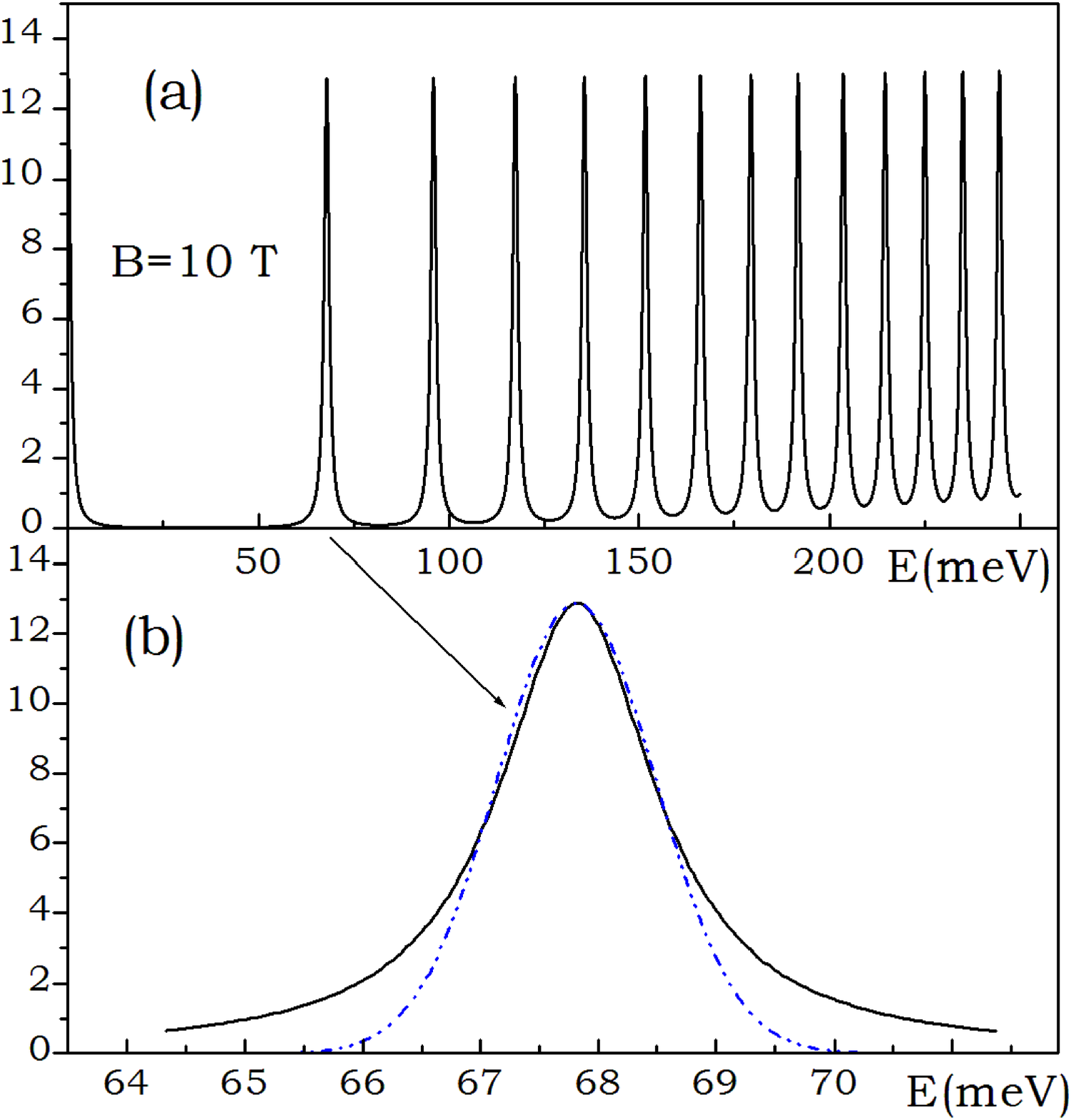}
                 \caption{a) Positive energy branch of density of states obtained from
                 the regularized GF for monolayer graphene at a magnetic field~$B=10$~T.
                 Positions of peaks are given by~$E=\hbar\Omega\sqrt{n}$
                 with~$\Omega=67.82$~meV and~$n=0,\pm 1, \pm 2 \ldots$.
                 The broadening parameter is~$\eta=1$~meV~\cite{Jiang2007,Yang2010}.
                 b) Detailed plot of DOS peak for the Landau level~$n=1$.
                 The difference between exact peak form and the Lorentz function
                 is negligible within the plot accuracy.
                 Dotted line: Gaussian like profile of the same FWHM.} \label{FigMLGPeak}
\end{figure}

In Fig.~\ref{FigMLGPeak}a we plot the unperturbed DOS as a
function of energy in monolayer graphene calculated
from the regular part of GF,
see Eqs.~(\ref{MLG_reg_uu}) and~(\ref{MLG_reg_ll}),
for~$B=10$~T and~$E\ge 0$. We assume a finite
Landau level width~$\eta=1$~meV, which agrees with experimental
estimations of~$\eta$ in graphene, see Refs.~\cite{Jiang2007,Yang2010}.
The DOS is a the series of nearly Lorentz-like peaks
centered at~$E_n=\hbar \Omega \sqrt{n}$,
with~$\Omega\simeq 67.82$~meV. For larger~$n$,
the peaks overlap with each other and DOS tends to a smooth function.
In Fig.~\ref{FigMLGPeak}b we
plot the DOS peak for Landau level~$n=1$.
We also plot the Gaussian-like peak having the same half-width.
Differences between the Gaussian and the Lorentz functions
exist in tails only, and they can be neglected for well-separated
peaks.

\begin{figure}
\includegraphics[width=8cm,height=8cm]{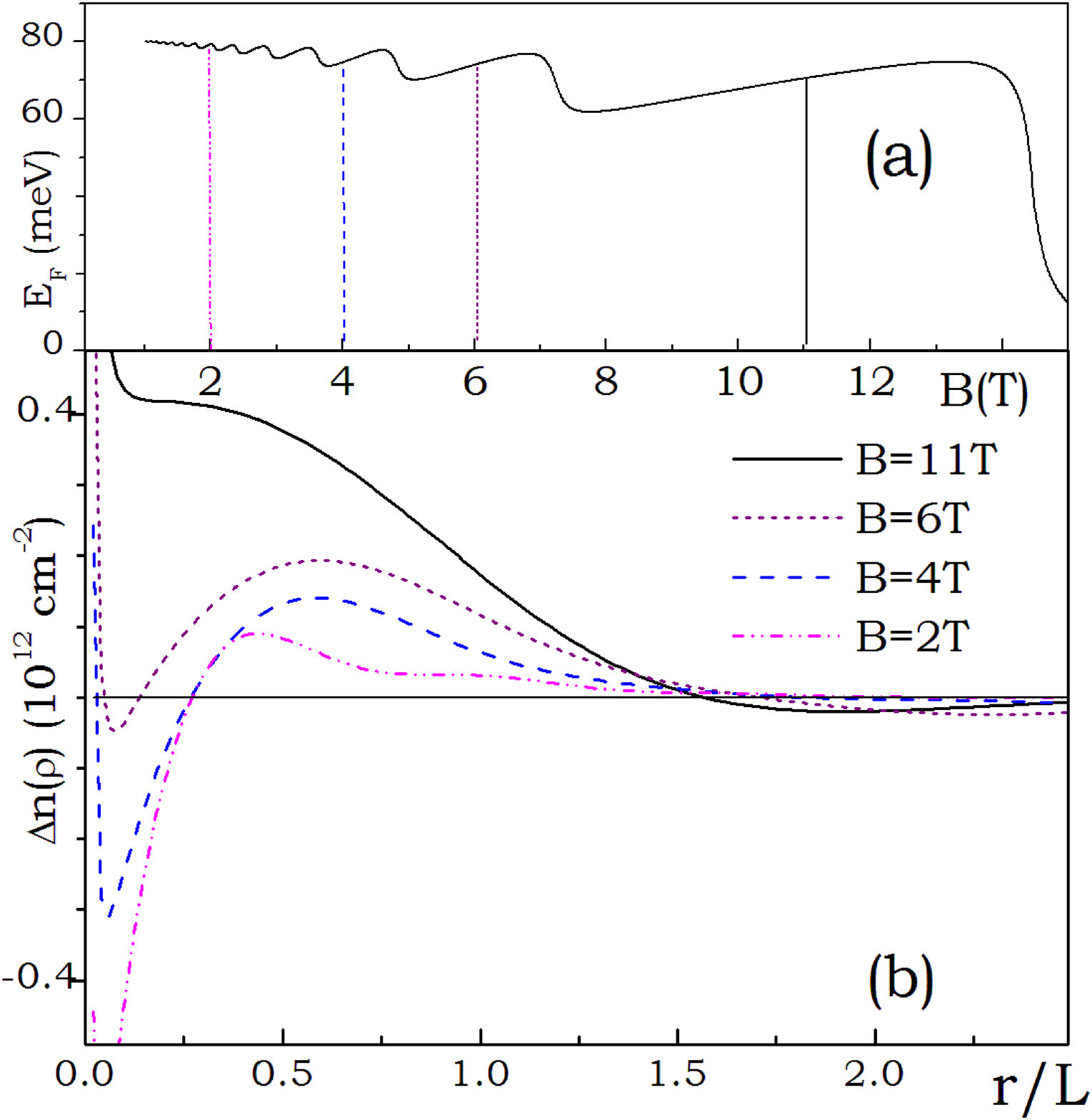}
                 \caption{a) Position of the Fermi level for monolayer
                 graphene in a magnetic field at constant electron
                 density~$n_e=0.35\times 10^{12}$~cm$^{-2}$~\cite{Ando2009}
                 and broadening~$\eta=1$~meV~\cite{Jiang2007,Yang2010}.
                 Vertical lines: magnetic fields
                 used in lower panel. b) Radial parts of induced electron
                 density~$\Delta n({\bm \rho})$
                 in monolayer graphene around delta-like neutral impurity in a magnetic
                 field. Parameter~$V_0=-56$~eV\AA$^2$ corresponds to nitrogen
                 impurity~\cite{Lambin2012}.
                 Distance is given in magnetic length~$L$ units.} \label{FigMLGvsB}
\end{figure}

\begin{figure}
\includegraphics[width=8cm,height=8cm]{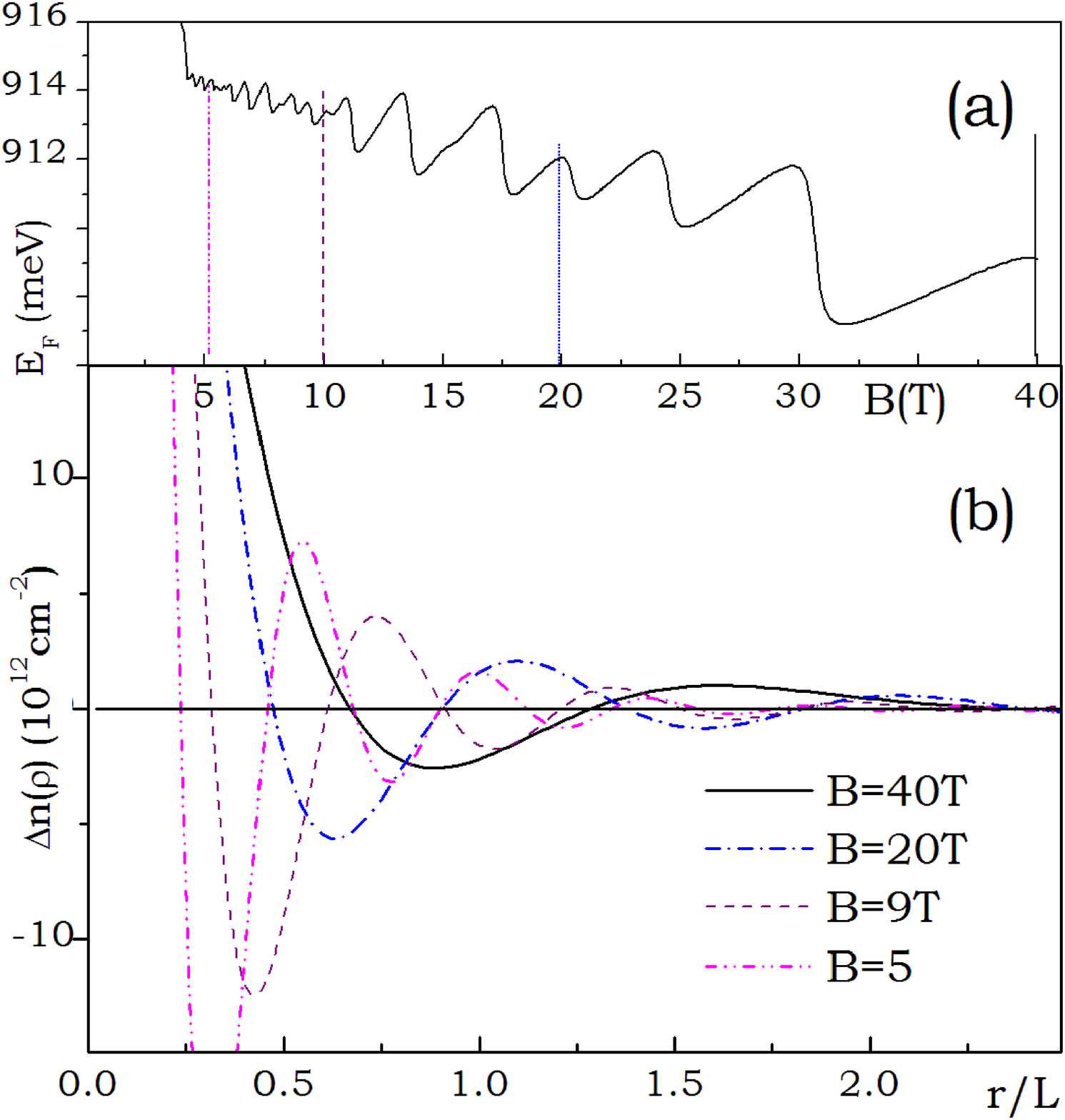}
                 \caption{a) Position of the Fermi level for WS$_2$ in a magnetic
                 field at constant electron density~$n_e=5.9\times 10^{12}$~cm$^{-2}$
                 and broadening~$\eta=0.74$~meV~\cite{Wang2017}.
                 $E_F$ is measured from the zero energy of the Hamiltonian~(\ref{VI_hH1B}), while
                 the bottoms of conduction
                 bands are at~$E_c=\Delta/2=895$~meV. Vertical lines: magnetic fields
                 used in lower panel. b) Radial parts of induced electron density~$\Delta n({\bm \rho})$
                 in WS$_2$ around delta-like neutral impurity in a magnetic field.
                 Parameter~$V_0=-4.3$~eV\AA$^2$ corresponds
                 to nickel impurity in high-T$_c$ superconductors~\cite{Wang2005}.
                 Distance is given in magnetic length~$L$ units.
                 Parabolic model of WS$_2$ bands is used.} \label{FigWS2vsB}
\end{figure}

As pointed out by Ando~\cite{Ando2009}, the concentrations of electrons ($n_e > 0$)
and of the holes ($n_e < 0$) in graphene can be controlled
over a wide range of concentrations:~$n_e \in [-5 \times 10^{13}...5 \times 10^{13}]$ cm$^{-2}$
by the gate voltage between graphene layer and heavily doped silicon.
Below we take~$n_e=0.35\times 10^{12}$ cm$^{-2}$, i.e. within the available range.
In Fig.~\ref{FigMLGvsB}a we plot the Fermi level as a function
of a magnetic field for monolayer graphene at~$T=0$ for a fixed electron concentration~$n_e$.
The vertical lines indicate magnetic fields for
which we calculate the oscillations of electron density.
The Fermi level oscillates with a magnetic field, c.f. Ref.~\cite{Zawadzki1984}.
The oscillations occur when the electrons
begin to fill the consecutive Landau level (LL),
since the 'capacity' of each LL increases with magnetic field.
For sufficiently large fields all electrons are in the lowest LL with~$n=0$,
and the Fermi level drops to energies around zero.

In Fig.~\ref{FigMLGvsB}b we plot the induced electron density calculated
from Eqs.~(\ref{hG_MLG_RR})
and~(\ref{Delta_n}) for a delta-like impurity placed at~${\bm \rho}_0={\bm 0}$ for
several values of magnetic field. We assume the impurity
potential~$V_0=-56$~eV\AA$^2$
which corresponds to the nitrogen impurity in graphene in
Ref.~\cite{Lambin2012}, see Discussion.
The calculations presented in Fig.~\ref{FigMLGvsB}b correspond to the
standard experimental configuration in which a sample is placed
in varying magnetic field. As seen in the figure, for all field
values the induced density oscillates with
the distance~$r$ from the impurity, and the density oscillations are
similar to FO in absence of fields.
The periods of oscillations become longer with increasing field
and the magnitudes of the oscillations increase.
For all values of magnetic field the oscillations disappear for~$r\ge 4L$.
For~$B=11$~T only one oscillation occurs, while for~$B=2$~T
the oscillations are irregular. To summarize, for typical
material parameters of graphene and a nitrogen impurity
it should be possible to observe FO in a magnetic field.

In Figs.~\ref{FigWS2vsB}a and~\ref{FigWS2vsB}b we present the results for~WS$_2$.
They are similar to those for monolayer graphene. The calculations
are performed using the parabolic model of energy bands, see Eq.~(\ref{g2D_PA}).
The material parameters for~WS$_2$ are listed in Table~1.
Following experimental results of Ref.~\cite{Wang2017} we take the electron
density~$n_e=5.9 \times 10^{12}$~cm$^{-2}$ and broadening of LL equal to 2~meV of FWHM.
This value corresponds to~$\eta=0.74$~meV for the Lorentz peak given
in Eq.~(\ref{g2D_reg_a}) in Appendix~C.
Bottoms of the two conduction bands are located at~$E_c=895$~meV,
and the Fermi energy is counted from~$E_c$.
We take~$V_0=-4.3$~eV$\AA^2$ which corresponds
to nickel impurity in high-T$_c$ superconductors~\cite{Wang2005},
see Discussion.

The results in Fig.~\ref{FigWS2vsB}b are similar to these reported for graphene,
but the oscillations are more pronounced. For all values of
magnetic field one observes a few spatial oscillations vanishing for~$r\ge 3L$.
The oscillation periods increase with the field, but their amplitudes are less sensitive
to field values. Since here~$V_0 <0$, the induced electron
density increases in the vicinity of impurity because electrons are attracted
by the well potential.
Note that, for some~$r$, the induced electron density~$\Delta n({\bm \rho})$ may
exceed the density of free electron gas~$n_e=5.9\times 10^{12}$~cm$^{-2}$.
This is an artifact of our model resulting from disregarding the many-body
interactions in the electron gas, see Discussion.

In Figs.~\ref{FigMLGvsB}a and~\ref{FigWS2vsB}a the Fermi levels in the two
samples exhibit saw-like
oscillations with increasing magnetic field. To explain this behavior let us
recall that the degeneracy of one LL
is:~$n_{LL} = eB/h \simeq 2.4 \times$ 10$^{10}\ B$ [cm$^{-2}$],
where~$B$ is measured in Tesla. For fixed electron density~$n_e$,
the increase of magnetic field results
in pushing electrons towards lower Landau numbers~$n$~\cite{Zawadzki1984}.
Finally, for sufficiently high fields there is~$n_{LL=0} > n_e$
and all electrons are at the lowest level.
The oscillations in Figs.~\ref{FigMLGvsB}a and~\ref{FigWS2vsB}a
are smoothed because of finite level widths~$\Gamma$.

\begin{figure}
\includegraphics[width=8cm,height=8cm]{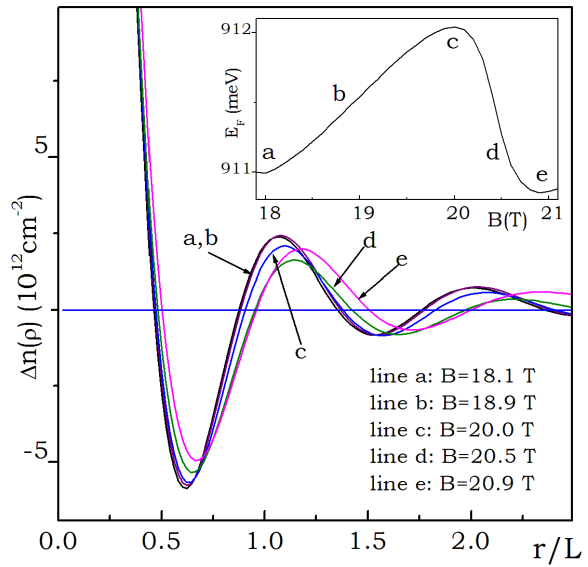}
                 \caption{
                 Radial parts of induced electron density~$\Delta n({\bm \rho})$
                 in WS$_2$ around delta-like neutral impurity in a magnetic field.
                 Model parameters are:~$V_0=-4.3$~eV\AA$^2$
                 and~$\eta=0.74$~meV, see Fig.~\ref{FigWS2vsB}.
                 Inset: The same as in Fig.~\ref{FigWS2vsB}a but
                 for one peak of~$E_F$ oscillations. Letters~$a,b,c,d,e$
                 indicate magnetic fields used in the main figure.} \label{FigWS2EfPeak}
\end{figure}

In two previous pictures we analyzed the FO in a wide scale of magnetic fields. Now we
concentrate on FO in WS$_2$ for varying occupation of single Landau level
close to~$B=20$~T (see Fig.~\ref{FigWS2vsB}a),
which corresponds to~$n=3$. In Fig.~\ref{FigWS2EfPeak} we calculate FO for five
LL occupations and the offset we indicate magnetic fields
corresponding to the five cases.
General conclusion from Fig.~\ref{FigWS2EfPeak} is that FO weakly depend on
degree of occupation.

\begin{figure}
\includegraphics[width=8cm,height=8cm]{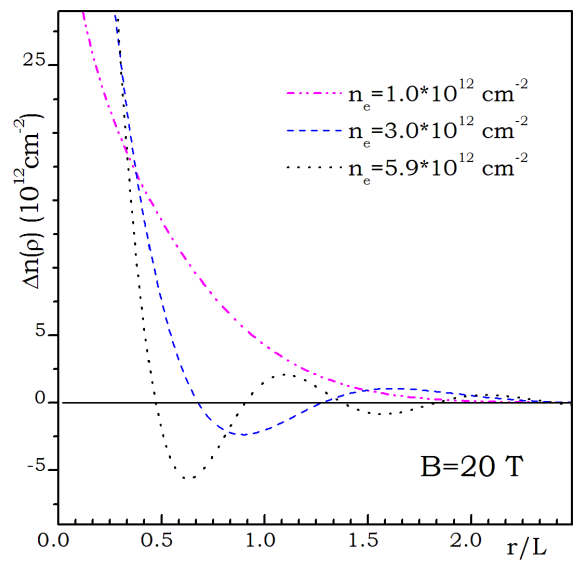}
                 \caption{Radial parts of induced electron density~$\Delta n({\bm \rho})$ in WS$_2$
                  around delta-like neutral impurity for a constant
                  magnetic field~$B=20$~T for various concentrations~$n_e$
                  of electron gas. Model parameters are:~$V_0=-4.3$~eV\AA$^2$
                  and~$\eta=0.74$~meV, see Fig.~\ref{FigWS2vsB}} \label{FigWS2vsNe}
\end{figure}

In Fig.~\ref{FigWS2vsNe} we show results for
several samples of different electron concentrations~$n_e$ in the same
magnetic field of~$B=20$~T. For high~$n_e$, the induced electron density oscillates,
while for low~$n_e$ the oscillations vanish. In all cases the oscillations
disappear for~$r \ge 3L$. For low~$n_e$ the Fermi level is located in
the lowest Landau level~$n=0$ (in the so called quantum strong field limit)
and the induced electron density decays exponentially with~$r$ with a characteristic
length on the order of~$L$. Numerical fit in Fig.~\ref{FigWS2vsNe} to the
curve for~$n_e=10^{12}$ cm$^{-2}$ gives, to a high
accuracy, the Gaussian form of the decay:~$\Delta n({\bm \rho}) \propto A\exp[-a(r-r_0)^2]$,
where~$A,a,r_0>0$ are three positive constants.
This result is similar to predictions for 3D
electron gas and for delta-like and screened Coulomb-like
impurities~\cite{Rensink1968,Horing1969a}.

\begin{figure}
\includegraphics[width=8cm,height=8cm]{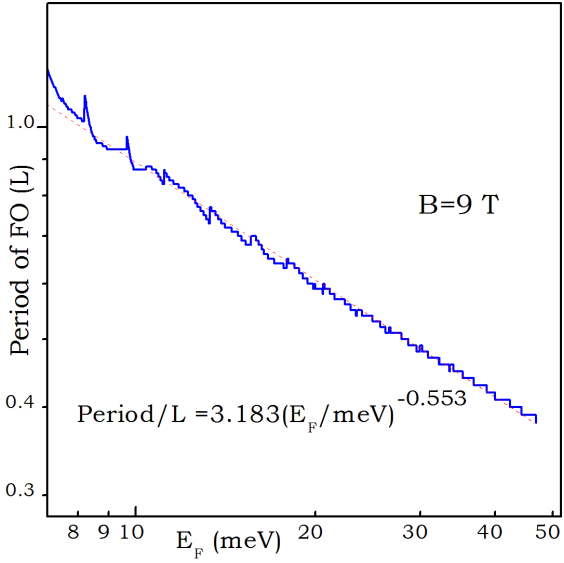}
                 \caption{Period of Friedel oscillations vs. Fermi energy in WS$_2$
                 in a magnetic field~$B=9$~T. Model parameters are:~$V_0=-4.3$~eV\AA$^2$
                 and~$\eta=0.74$~meV, see Fig.~\ref{FigWS2vsB}.} \label{FigWS2Period}
\end{figure}

In the field-free case FO arise from the sharp change
of electron density in the~$k$ space in the vicinity of the Fermi sphere,
which leads to spatial oscillations of the density with the
period~$T_{FO} \propto \pi/k_F$, where~$k_F$ is the
Fermi vector. For free electrons in 2D there is~$k_F\propto E_F^{1/2}$,
which gives
\begin{equation} \label{TFO}
 T_{FO} \propto \frac{1}{\sqrt{E_F}}.
\end{equation}
The main physical difference between 2D electrons in the presence or
absence of a magnetic field is, that in the former case, the electron motion
is fully quantized within the plane perpendicular to the field and for
nonzero field there is no~$k_F$ vector. However, since the Fermi energy is a well
defined quantity both in
the presence and absence of the field, we may expect that Eq.~(\ref{TFO})
remains valid also for the FO at nonzero magnetic field.

To verify this expectation, we plot in Fig.~\ref{FigWS2Period} the period
of FO as a function of~$E_F$ for~$B=9$~T.
As before,~$E_F$ is measured from the conduction bands edges.
The oscillation periods~$T_{FO}$ are computed numerically
as an average distance between several consecutive minima and maxima of
induced electron density.
The results are plotted in the logarithmic scale.
As seen in Fig.~\ref{FigWS2Period}, the periods~$T_{FO}$
follow the formula of Eq.~(\ref{TFO}) to a high accuracy.

Equation~(\ref{TFO}) qualitatively explains the non-oscillating
behavior of induced density in the quantum strong field limit.
For group-VI dichalogenides, the lowest energy level in the conduction band
is at~$E=\Delta/2$ and in the quantum strong field limit there
is~$E_F=0$, since~$E_F$ is measured from the bottom of conduction band,
see Section~II. Then the period of oscillations in Eq.~(\ref{TFO})
is infinite, which is equivalent to the non-oscillating decay
of induced electron density.

\begin{figure}
\includegraphics[width=8cm,height=8cm]{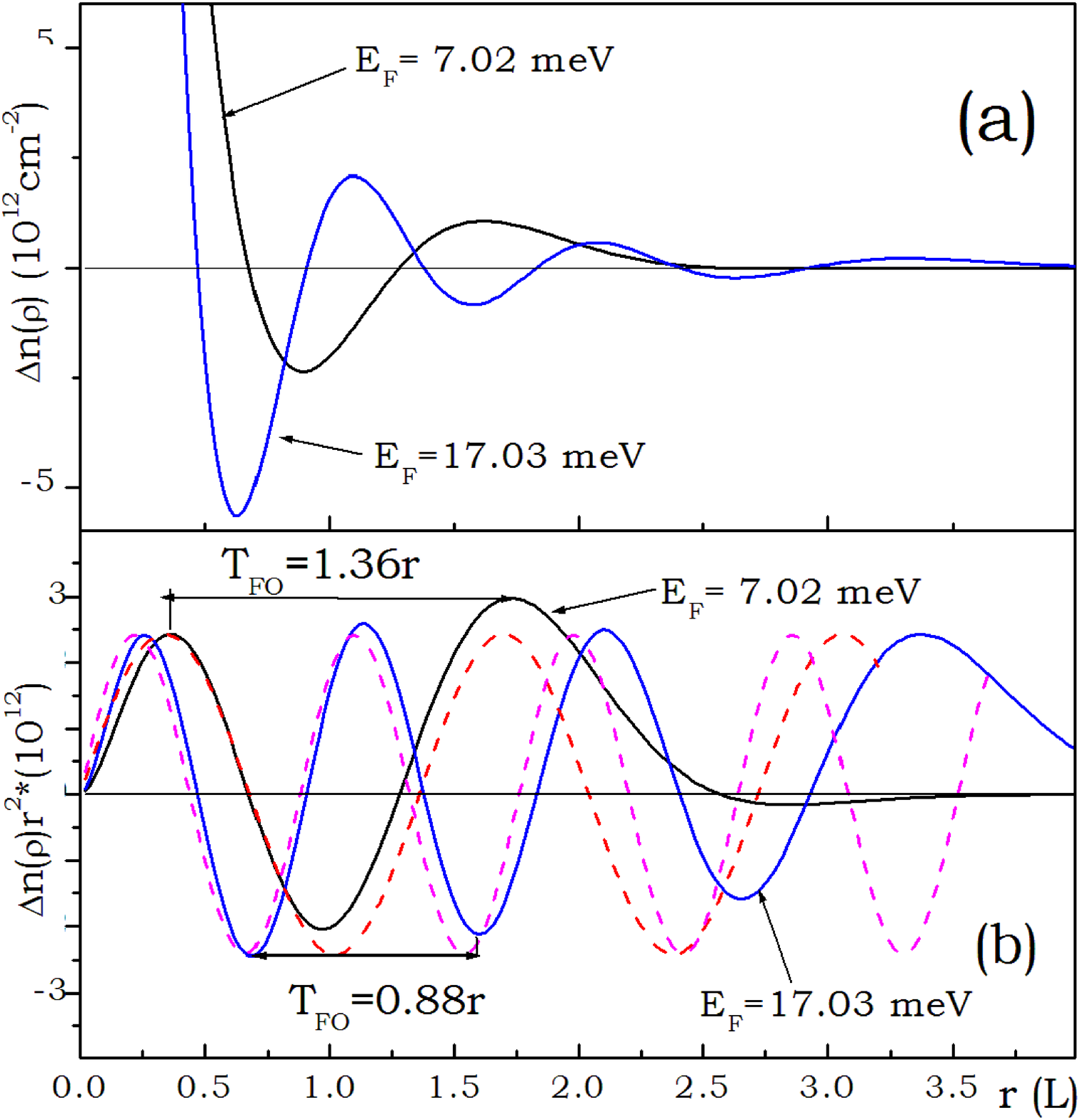}
\caption{a) Radial parts of induced electron density~$\Delta n({\bm \rho})$ in WS$_2$ around
        delta-like impurity for~$n_e=5.9\times 10^{12}$~cm$^{-2}$
        and~$n_e=3.0\times 10^{12}$~cm$^{-2}$. The magnetic field is~$B=20$~T.
        Fermi energies are located at~$E_F=17.03$~meV and~$E_F=7.02$~meV above
        the bottom of conduction band, respectively.
        Model parameters are:~$V_0=-4.3$~eV\AA$^2$
        and~$\eta=0.74$~meV, see Fig.~\ref{FigWS2vsB}.
        b) Solid lines: results from upper panel multiplied by~$r^2$.
        Dotted lines: functions~$\sin(2\pi r/T_{FO})$ for two~$T_{FO}$ values.}
        \label{FigWS2vsBr2}
\end{figure}

As predicted by Rensink~\cite{Rensink1968} and Horing~\cite{Horing1969a},
the FO in 3D should be expressed in terms of the
sine or cosine functions of~$(2k_Fr)$ divided by~$r^{-3}$, where~$k_F$ is the Fermi vector.
Having in mind this result and taking Eq.~(\ref{TFO}) we propose the following approximate
formula for FO in the parabolic band model for
electrons in group-VI dichalogenides
\begin{equation} \label{Dn_guess}
 \Delta n({\bm \rho}) \simeq \delta n \frac{\sin(2\pi r /T_{FO})}{r^2},
\end{equation}
where~$T_{FO}$ is given in Eq.~(\ref{TFO}) and~$\delta n$ is a constant.

To verify the validity of Eq.~(\ref{Dn_guess}), we plot in
Fig.~\ref{FigWS2vsBr2}a the
induced electron density~$\Delta n({\bm \rho})$ in a magnetic field~$B=20$~T
for two values of electron concentration~$n_e$ which
correspond to two values of the Fermi energy.
In Fig.~\ref{FigWS2vsBr2}b we isolate the oscillating part of
the induced density:~$\Delta n({\bm \rho}) \times r^2$.
It is seen that both solid curves oscillate with finite and nearly
constant amplitudes. This leads to conclusion
that, for large~$r$, FO in a magnetic field
decay as~$r^{-2}$. This agrees with the field-free result in Eq.~(\ref{Friedel_D}).
As follows from Eq.~(\ref{TFO}), for a fixed magnetic field the product~$T_{FO}\sqrt{E_F}$
is a constant, and this result is obtained for the two solid
lines in Fig.~\ref{FigWS2vsBr2}b.
Since the first oscillations of solid lines in Fig.~\ref{FigWS2vsBr2}b
resemble the sines functions we plot in Fig.~\ref{FigWS2vsBr2}b two
sine functions given by the numerator in Eq.~(\ref{Dn_guess}),
with the periods shown in the figure. For the first few cycles there is
a good agreement between the exact results and the sine functions.

The results presented in Fig.~\ref{FigWS2vsBr2}b and
in Eqs.~(\ref{TFO}) and~(\ref{Dn_guess})
are extensions of the predictions of Refs.~\cite{Rensink1968,Horing1969a}
concerning 2D electron gas in materials with honeycomb lattice.
They are consistent with the perturbation approach,
but our results are valid (within the validity of non-interacting gas model)
for arbitrary electron concentrations, magnetic fields, impurity potentials~$V_0$ and
distances from the impurity, while the results of Refs.~\cite{Rensink1968,Horing1969a}
have some limitations, see Discussion.

\begin{figure}
\includegraphics[width=8cm,height=8cm]{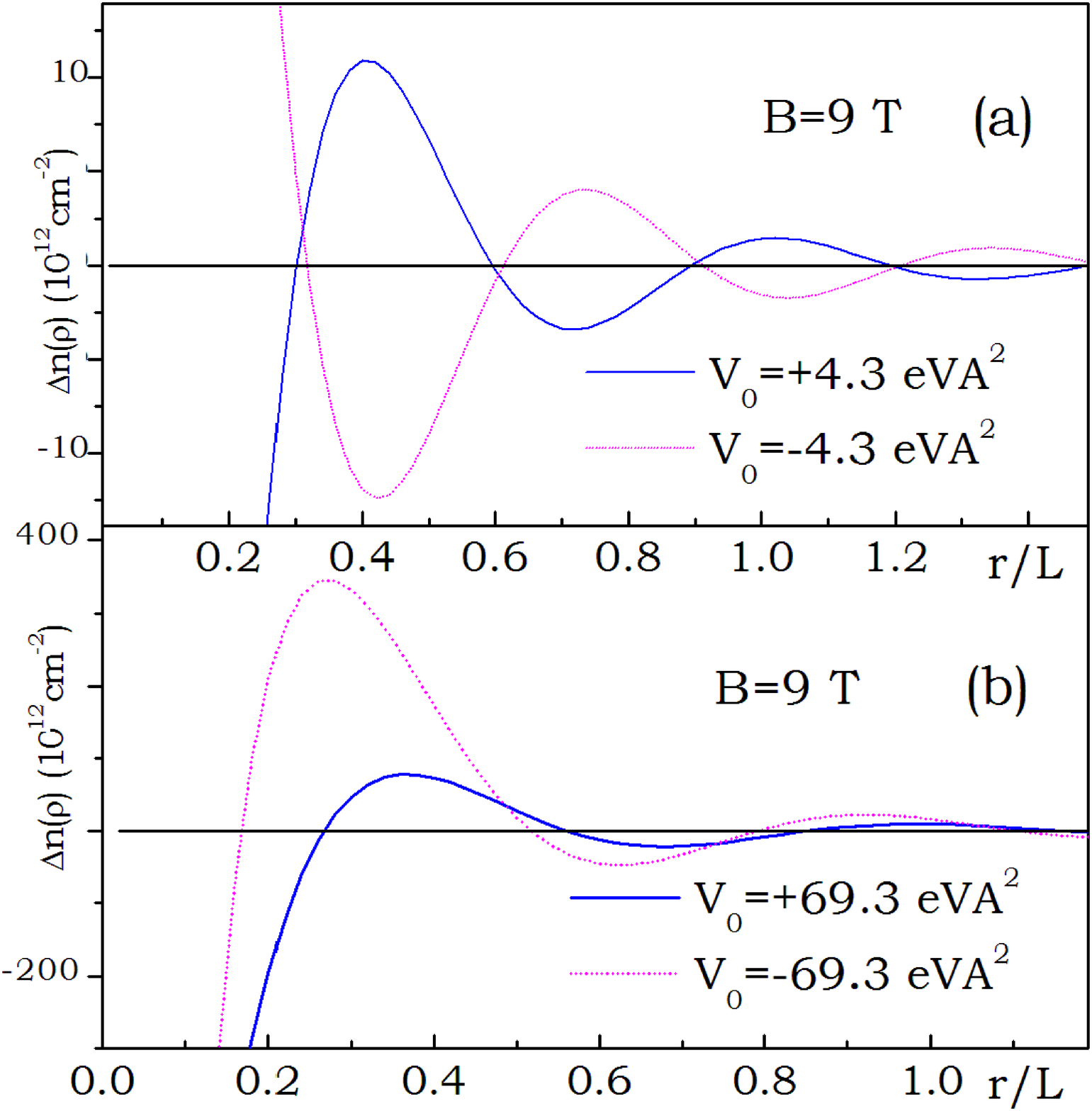}
                 \caption{Radial part of induced electron density~$\Delta n({\bm \rho})$ in WS$_2$
                 around delta-like neutral impurity in a magnetic field
                 for attractive and repulsive~$V_0$. a) For small~$|V_0|$ values
                 there is~$\Delta n({\bm \rho})~\propto V_0$, see Eq.~(\ref{hG_smV}).
                 b) For large~$|V_0|$ values~$\Delta n({\bm \rho})$ weakly depends
                 on~$V_0$, see Eq.~(\ref{hG_laV}).}
                 \label{FigWSvsV0}
\end{figure}

In Fig.~\ref{FigWSvsV0} we consider induced electron density~$\Delta n({\bm \rho})$
for small and large values of~$|V_0|$.
For small~$|V_0|$ one can neglect the denominators in Eq.~(\ref{hG_2D_RR}) and obtain in
the lowest order of the Born series
\begin{equation} \label{hG_smV}
 \hG_{\bm 1,\bm 2} \simeq \hg_{\bm 1,\bm 2} +
 \hg_{\bm 1,\bm 0} V_r \hg_{\bm 0,\bm 2} + \ldots.
\end{equation}
Taking~$V_r \simeq V_0$ we find that, for small~$V_0$, there is~$\Delta n({\bm \rho})\propto V_0$.
To show this we plot~$\Delta n({\bm \rho})$ for two opposite~$V_0$ values and obtain two
symmetric lines.

In Fig.~\ref{FigWSvsV0}b we plot~$\Delta n({\bm \rho})$ for large~$|V_0|$.
In this case one can neglect the unity
in the denominator of Eq.~(\ref{hG_2D_RR}) which gives
\begin{equation} \label{hG_laV}
 \hG_{\bm 1,\bm 2} \simeq \hg_{\bm 1,\bm 2} + \hg_{\bm 1,\bm 0}
 \frac{1}{\hg^{reg}} \hg_{\bm 0,\bm 2},
\end{equation}
i.e. in this limit GF of the system and the resulting~$\Delta n(\bm \rho)$
do not depend on the impurity potential having an universal character.

\section{Discussion}

As shown in Figs.~\ref{FigWS2Period} and~\ref{FigWS2vsBr2},
equations~(\ref{TFO}) and~(\ref{Dn_guess}) are valid for
the ideal 2D electron gas and for that in group-VI dichalogenides
in the parabolic approximation of energy band for typical material parameters.
Our calculations for monolayer graphene and WS$_2$ in the nonparabolic bands
model suggest that Eq.~(\ref{Dn_guess}) should be replaced by
\begin{equation} \label{Dn_beyond}
\Delta n({\bm \rho})
 \simeq \delta n \frac{F_{osc}[r,T_{osc}(r)]}{r^{2+\alpha}},
\end{equation}
in which~$0< \alpha < 0.2$ and~$F_{osc}$ is a bound and oscillating
function of the distance. This results agrees with predictions
for 2D massless Dirac fermions reported in Ref.~\cite{Thakur2017}, where
for large~$\rho$ the authors predicted~$\Delta n({\bm \rho}) \propto\rho^{-3}$.
In Eq.~(\ref{Dn_beyond}) the oscillation period $T_{osc}(r)$
increases with~$r$ in a similar way to that
shown in Fig.~\ref{FigWS2vsBr2}b.  All quantities entering Eq.~(\ref{Dn_beyond})
depend on~$V_0$. The numerical algorithms given in Appendix~B allow one
to calculate~$\Delta n({\bm \rho})$ for various material parameters.

We considered a delta-like impurity potential of neutral impurity.
With this choice it is possible to sum the Born series and obtain
the exact GF. To estimate
the range of validity of this model we first assume that a typical
impurity size is on the order~$r_i\approx 1-2$~\AA, i.e. the size of
an atom in the lattice. The impurity can be treated as delta-like if its
size is much smaller than the oscillation period~$T_{FO}$.
As seen in Fig.~\ref{FigWS2Period}, for a fixed magnetic field~$T_{FO}$
decreases with the Fermi energy and electron concentration~$n_e$.
For parameters in Fig.~\ref{FigWS2Period} our model is valid for~$T_{FO}\ge 20\AA$,
i.e. for all~$E_F$ values shown in the figure.

To consider the physical sense of approximating an impurity potential
by the delta function we assume the potential to be short-range with a characteristic
length~$a$ centered at~${\bm \rho}=0$. Then the Dyson equation
for GF is, see Eq.~(\ref{hG_2D_0})
\begin{eqnarray}
 \label{hG_2D_V0}
 \hG_{\bm 1,\bm 2} &=& \hg_{\bm 1,\bm 2} + \int_{|\bm \rho|< a} \hg_{\bm 1,\bm 3}
 V({\bm \rho}_3)\hG_{\bm 3,\bm 2} d^2{\bm \rho}_3 \\
 \label{hG_2D_bV0}
 &\simeq & \hg_{\bm 1,\bm 2} + \hg_{\bm 1,\bm 0} V_0 \hG_{\bm 0,\bm 2},
\end{eqnarray}
in which
\begin{equation} \label{Int_V0}
 V_0 = \int_{|{\bm \rho}_3|< a} V({\bm \rho}_3) d^2{\bm \rho}_3 \simeq
 V(\bm 0)a^2 \hspace*{0.25em} {\rm [J \times m^2]} .
\end{equation}
The model of a delta-like potential is valid if the integral in Eq.~(\ref{hG_2D_V0})
is well approximated by Eqs.~(\ref{hG_2D_bV0}) and~(\ref{Int_V0}).
In realistic systems the delta-like potential describes a potential
of neutral impurity in the lattice, potential of a vacancy,
contact potential arising from electron-nuclei interactions or
potential of a screened ion having a short screening length.
Equation~(\ref{Int_V0}) explains the physical units of~$V_0$ entering
into the Dyson equation, namely~J$\times$m$^2$.

Now we discuss the magnitudes of~$V_0$ used in the calculations
is Section~III. To select~$V_0$ we take results reported in the literature
concerning impurities, vacancies, dislocations or potential drops at surfaces.
Wang {\it et al.}~\cite{Wang2005} reported:~$V_0=-4.3$~eV$\AA^2$ for nickel
impurity,~$V_0=11.2$~eV$\AA^2$
for zinc impurity and~$V_0=69.3$~eV$\AA^2$ for a vacancy, respectively.
These estimations are based on ab-initio calculations of~Cu panels
in high-T$_c$\ superconductors.
Lang and Kohn calculated the effective one-electron potential
for metal surface in the jellium model, (see Fig.~3 of
Ref.~\cite{Lang1970} or Fig.~2 of Ref.~\cite{Dobson1995}),
which gives~$V_0 =16$~eV\AA$^2$.

Finally, one can estimate~$V_0$ from impurity potentials used in ab-initio calculations.
In Ref.~\cite{Lambin2012}, the potential of nitrogen impurity in graphene
was assumed in the Gaussian form
\begin{equation} \label{V0_Nitro}
 V(\bm \rho)= \epsilon_C -4.0\exp\left(-\frac{\rho^2}{2\sigma^2}\right)
 \hspace*{0.5em} {\rm [eV]},
\end{equation}
where~$\epsilon_C$ is the asymptotic bulk value of the on-site parameter of carbon,
and~$\sigma=1.5$~\AA. Performing integration in Eq.~(\ref{Int_V0})
for potential in Eq.~(\ref{V0_Nitro})
we find:~$V_0=-2\pi\times 4.0\times \sigma^2=-56$~eV\AA$^2$.

\begin{figure}
\includegraphics[width=8cm,height=8cm]{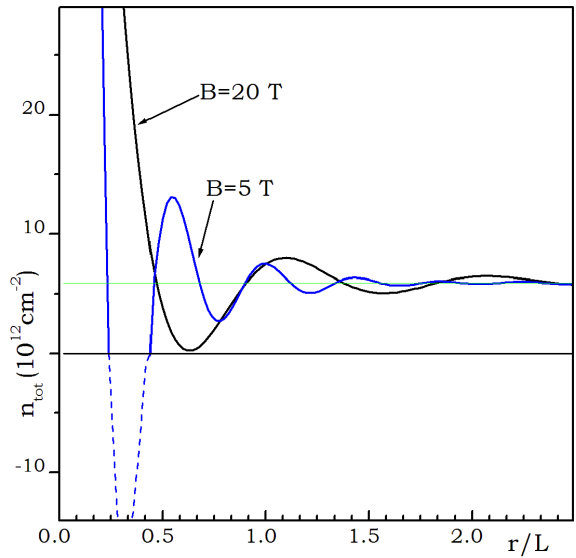}
                 \caption{Radial parts of total electron
                 density~$n_{tot}=n_e + \Delta n(\bm \rho)$
                 calculated for WS$_2$ around delta-like neutral impurity
                 in a magnetic field. Dotted line indicates schematically
                 non-physical negative total electron density.}
                 \label{FigWS2vsBNL}
\end{figure}

As seen in Fig.~\ref{FigMLGvsB}, for monolayer graphene
the induced electron density does not exceed
the background density~$n_e=0.35\times 10^{12}$~cm$^{-2}$, so that the total
density of electron gas:~$n_{tot} = n_e+\Delta n(\bm {\rho}) > 0$ for all~${\bm \rho}$.
However, in WS$_2$, for small~$r$ and low magnetic fields the magnitude
of~$\Delta n(\bm {\rho})$ exceeds the electron density~$n_e=5.9\times 10^{12}$~cm$^{-2}$.
Thus, for some~$r$, there is~$n_{tot}<0$, which is an artifact of the model.

To analyze this question we note first that the calculations in Section~II are
exact in the sense that we sum the perturbation series in Eq.~(\ref{hG_2D_Ex})
to the infinite order, so that our results are valid for
arbitrary~$V_0$ and~$E$ values. However, there remain limitations of the
free-electron model. To estimate the validity of the free-electron approximation,
we plot schematically in Fig.~\ref{FigWS2vsBNL} the total electron density
with an additional condition~$n_{tot} \geq 0$. The dotted line indicates the non-physical
case of negative~$n_{tot}$. It is seen that, for~$B=20$~T, for all~$r$
there is~$n_{tot} >0$, and the model of noninteracting electron gas
used above is correct. Similarly, for~$B=5$~T there is~$n_{tot} >0$
for~$r>0.5 L$, which determines the range of validity of the model.
The conclusion from Fig.~\ref{FigWS2vsBNL} is that the model
of non-interacting electrons applied to the calculation of FO
in a magnetic field for electrons in 2D materials with a honeycomb lattice
is valid under certain conditions. First, the impurity
potential~$V_0$ should be weak and possibly attractive ($V_0<0$).
Next, both the magnetic field and electron density~$n_e$ should be high.
Finally, in some cases the results for small~$r$ may be doubtful,
but they become reasonable for large distances from the impurity.

In the literature describing the electron gas in metals there exist
quantities obtained by the theory of non-interacting particles which are unphysical
while those obtained within the many-body
theories do not suffer from this deficiency. For example the so-called pair
distribution function~$g(r)$, describing
the probability of finding an electron in a distance~$r$
from the impurity, should be strictly
positive. However, as pointed out in Ref.~\cite{Sjolander1972},
the calculations of~$g(r)$ within linear model lead for some~$r$ to negative
values of~$g(r)$. On the other hand, by including electron-electron interactions,
exchange and non-linear effects one obtains positive values of~$g(r)$
for all distances, see Refs.~\cite{Perdew1992,Vericat2011}.
For this reason we expect that the account of the electron-electron interaction would correct the
nonphysical effects for FO oscillations at small distances from the impurity.

Let us now turn to the regularization procedure and the divergences
in~$\hg_{\bm 0,\bm 0}$. The origin of these divergencies arises
from the divergences of harmonic series in Eq.~(\ref{g2D_r0}) for large~$n$, i.e.
for large energies. However, the~$\bm k \cdot \bm p$ model is valid for
energies near band extrema, i.e. for small~$n$. The divergences
appearing in Eq.~(\ref{g2D_r0}) are {\it artifacts} of the model
which justifies the use of regularization procedure described in Section~II.

As mentioned in Introduction, the regularization procedure
is based on two elements. First, for all models in Section~II
there exist analytical expressions for sums over Laguerre polynomials
see Eq.~(\ref{g2D_Ex}). Second, knowing the limit of the Whittaker
function for small arguments one can isolate the divergent
and regular parts of GF,
see Eqs.~(\ref{Whitt_k0})--(\ref{g2D_reg}).
If one were unable to sum up an infinite series in Eq.~(\ref{g2D_1}),
one would be unable to isolate divergent and regular parts of~$\hg$.
Therefore, the knowledge of analytical form of the one-electron
GF is necessary to perform the
regularization procedure described in Section~II.

There exists another consequence of the divergencies in~$\hg_{\bm 0,\bm 0}$
of Eq.~(\ref{hG_2D_Ex}), namely the impossibility of the perturbation expansion
of the Dyson equation into the Born-von Neumann series.
Treating, incorrectly,~$V_0\hg_{\bm 0,\bm 0}$ as an expansion parameter, one obtains
\[\hG_{\bm 1,\bm 2} \simeq \hg_{\bm 1,\bm 2} + \hg_{\bm 1,\bm 0}V_0 \hg_{\bm 0,\bm 2} +
 \hg_{\bm 1,\bm 0}\left[V_0\hg_{\bm 0,\bm 0}V_0 \right] \hg_{\bm 0,\bm 2} + \ldots .\]
In the above series the first and second terms are finite, but
the remaining terms diverge because they
include the powers of~$\hg_{\bm 0,\bm 0}$. There is no rigorous method of
removing these divergencies from the above series.
The correct way of obtaining the perturbation series of Born-von Neumann
is the use of Eq.~(\ref{hG_2D_RR}) instead Eq.~(\ref{hG_2D_Ex}):
\begin{equation} \label{hG_BornS}
 \hG_{\bm 1,\bm 2} \simeq \hg_{\bm 1,\bm 2} + \hg_{\bm 1,\bm 0}V_r \hg_{\bm 0,\bm 2} +
 \hg_{\bm 1,\bm 0}\left[V_r\hg_{\bm 0,\bm 0}^{reg}V_r \right] \hg_{\bm 0,\bm 2} + \ldots,
\end{equation}
where~$V_r$ is defined in Eq.~(\ref{V_RR}). The above expansion holds
for~$y=V_r\hg_{\bm 0,\bm 0} \ll 1$.

In Eq.~(\ref{V_RR}) we defined the regularized potential~$V_r$ and claimed
that, in practice, there is~$V_r\simeq V_0$. Now we discuss
validity of this assumption. Let us consider monolayer graphene for~$B=10$~T.
In Eq.~(\ref{hG_MLG_RR})
we take:~$\hbar\Omega=67.8$~meV,~$L=81.1$~\AA\ and~$V_0=-56$~eV\AA$^2$.
For~$\zeta=2.7$~\AA, i.e. for the size of a neutral atom in the lattice we
obtain from Eq.~(\ref{V_RR}):~$|V_0\hg_{\bm 0,\bm 0}| = 0.04$,
and one can approximate~$V_r\simeq V_0$.
For~$\zeta = 5$~fm, i.e. the size of a typical atomic nuclei,
one obtains:~$|V_0\hg_{\bm 0,\bm 0}| = 0.39$, which gives~$V_r=1.64 V_0$
for negative~$V_0$ and~$V_r=0.72 V_0$ for positive~$V_0$.
Then there is~$V_r=cV_0$ where~$c$ is still on the order of unity.
The potential of atomic nuclei is the narrowest realistic potential
appearing in the solid state materials, so that the regularization
procedure in Section~II leads to reasonable results.

There exist several other approaches allowing one
to overcome the problem of divergence of~$\hg_{\bm 0,\bm 0}$
in Eq.~(\ref{hG_2D_Ex}).
First, one can truncate the harmonic series in Eq.~(\ref{g2D_r0})
at certain~$n_{max}$, e.g. for energies~$\hbar\Omega n_{max}$ exceeding
the electron energies in the first BZ. Second, one can add a
convergence factor to the series in Eq.~(\ref{g2D_r0}),
allowing one to sum up the series.
Finally, one can replace the~${\bm k} \cdot {\bm p}$ Hamiltonian
by the more accurate tight-binding model, which automatically introduces
a cut-off in the energy scale.
Each of the above methods handles the problem of divergence, but it
introduces either an artificial cut-off or convergence factors,
or leads to a non-analytical form of the one-electron GF.
The regularization procedure described in Section~II is not better
than the discussed alternatives, but is more elegant and rigorous.

In our approach we assume low concentrations of neutral impurities.
This assumption is valid when the average distance~$R_{NN}$ between
nearest-neighbor impurities exceeds the range of density oscillations.
As seen in Figs.~\ref{FigMLGvsB} and~\ref{FigWS2vsB},
there is~$R_{NN} >3 L$, depending on the
strength of a magnetic field, which gives~$R_{NN}> 200$~\AA.
This corresponds to impurity concentrations~$n_i < 0.25\times 10^{12}$~cm$^{-2}$.
For higher~$n_i$, the densities induced by the two neighboring impurities
overlap with each other and the correlations effects between the two
impurities should be taken into account.

As mentioned above, position of the Fermi level is calculated for
impurity-free electron gas. This approach
is common in the literature, but it neglects the change of~$E_F$ due to presence of
impurities. The argument supporting this approach is that for low impurity concentration
the change of the electron gas density is spatial but not total.

Calculations of FO for 2D massive Dirac fermions in absence of magnetic field
using dielectric function approach were carried out by~\cite{Thakur2017}.
As to the Dirac fermions in a magnetic field, in Ref.~\cite{Yuan2012}
the authors calculated the polarization function for graphene in strong
fields, but they did not consider the density oscillations.

In our calculations we assumed~$T=0$ limit. The results can be generalized in the
standard way to finite temperatures by i) replacing the zero-temperature Green's
functions~$\hg(E)$ by~$\hg(i\hbar\omega_m)$, where~$\omega_m=(2\pi+1) k_BT/\hbar$
and~$m$ is an integer, ii) replacing in Eq.~(\ref{n_rho}) the integration over
the energy by the summation over~$m$, iii) adding the Fermi-Dirac distribution
function to the integral in Eq.~(\ref{n_rho}).
For finite~$T$ in absence of fields the electron distribution is spread over a wider
energy range than for~$T=0$ and there is a wider range of~${\bm k}$ vectors
allowed for the redistribution. As a result,
for~$T>0$ FO have the same period as for~$T=0$, but
there appear additional temperature-depending damping factors, see Ref.~\cite{Grassme1993}.
Similar effects are expected for FO in 2D electron gas in a magnetic field.

In the integral in Eq.~(\ref{n_rho}) we introduced the cut-off energy in the
lower limit which, in our calculations, is a few meV below the bottoms
of the conduction bands. We do not integrate over the filled valence bands
since they do not give contributions to the oscillating induced density.
However, we performed calculations treating the
cut-off energy~$E_x$ as a variable parameter. No significant change of the results
occurred, but there appeared oscillations having small amplitude
and frequency~$\omega_x =E_x/\hbar$.
To eliminate these artifacts, one should take~$E_x$ sufficiently
deep in the valence bands.

Promising experimental methods for observation of FO
in 2D materials with honeycomb lattice are the scanning
tunneling spectroscopy and the scanning tunneling microscopy, both
successfully used for observation of the FO in absence
of fields~\cite{Kanisawa2001,Hasegawa2007,Sessi2015}.
In these experiments, usually performed at low temperatures,
the modulation of the local density of states can be
resolved by differential conductivity maps~($dI/dU$).
Recently, similar measurements were performed by
Misra {\it et al.} with the scanning tunneling microscope
at high magnetic field~\cite{Misra2013}.
This method seems to be appropriate for experimental observation
of FO in a magnetic field for 2D electron gases.

\section{Summary}
The Friedel density oscillations induced by a delta-like neutral impurity in 2D electron
gases in the presence of a magnetic field are calculated.
Exact renormalized Green's functions obtained by an exact summation of the corresponding Dyson
series are used in the calculations. The renormalization procedures are first demonstrated
using the simple case of free 2D electron gas and then the developed methods are used to treat
the realistic cases of monolayer graphene and group-VI dicalchogenides employing
the appropriate band structures of these materials. Final results for FO are presented
which are valid for wide ranges of impurity potential strengths, electron densities,
magnetic fields and distances from impurity. Realistic models of neutral impurities
in the materials of interest are employed. It is found that, for weak impurity potentials,
the FO amplitudes are proportional to the potential strength.
The obtained formulas for FO are discussed and compared to results for 3D electron
gases in a magnetic field. In particular, it is shown that the Fermi
vector in a 3D electron gas is replaced by a corresponding quantity
for a 2D gas calculated from the Fermi energy.
In some particular situations: low magnetic fields, repulsive impurity potential
and small distances from impurity, the total calculated electron density becomes negative.
which indicates limitations of the one-body theory.

\appendix
\section{Green's function in group-VI dichalcogenides}

We present here details of calculations of the one-electron Green's function
for electrons in group-VI dichalcogenides in the presence of a magnetic field.
At the~${\bm K}$ point of BZ the eigenstates of the Hamiltonian~(\ref{VI_hH1B}) for~$n \geq 1$ are
\begin{eqnarray} \label{APsiP}
 \Psi_{nk_x+}({\bm \rho}) &=& \frac{e^{ik_xx}}{\sqrt{2\pi}{\cal N}_n} \left(\begin{array}{c}
 -\hbar\Omega\sqrt{n}\phi_{n-1}(\xi) \\ (E_n-E_w)\phi_{n}(\xi) \end{array}\right), \\
 \label{APsiM}
 \Psi_{nk_x-}({\bm \rho}) &=& \frac{e^{ik_xx}}{\sqrt{2\pi}{\cal N}_n} \left(\begin{array}{c}
 (E_n-E_w) \phi_{n-1}(\xi) \\ \hbar\Omega\sqrt{n}\phi_{n}(\xi) \end{array}\right),
\end{eqnarray}
and the eigenenergies~$E_{nk_x\epsilon}$ are given in Eq.~(\ref{VI_E1}).
There is~${\cal N}_n=\sqrt{2E_n(E_n-E_w)}$,
where~$E_w$ and~$E_n$ are defined in Eqs.~(\ref{VI_hH1}) and~(\ref{VI_E1}),
respectively. For~$n=0$ there is
\begin{eqnarray}
 \Psi_{0k_x+}({\bm \rho}) &=& 0 \\
 \Psi_{0k_x-}({\bm \rho}) &=& \frac{e^{ik_xx}}{\sqrt{2\pi}}
 \left(\begin{array}{c} 0 \\ \phi_0(\xi) \end{array}\right),
\end{eqnarray}
and the eigenenergy is~$E_{0k_x,-1,s_z}=-E_w+E_{s_z}$.
For further calculations it is convenient to introduce the
notation:~$\Psi_{nk_x+} = (\psi_+^u,\psi_+^l)^{\dagger}$
and~$\Psi_{nk_x-} = (\psi_-^u,\psi_-^l)^{\dagger}$.

Consider the Hamiltonian~(\ref{VI_hH1B})
with the eigenvalues given in Eq.~(\ref{VI_E1}) and eigenvectors
in Eqs.~(\ref{APsiP})--(\ref{APsiM}).
For~$\tau=+1$ and both~$s_z$ orientations the stationary GF of
the Hamiltonian~(\ref{VI_hH1B}) is
\begin{equation} \label{AG22}
 \hg_+({\bm \rho},{\bm \rho}',\cZ) =
 \left(\begin{array}{cc} \hga^{uu}_+ & \hga^{ul}_+ \\ \hga^{lu}_+ & \hga^{ll}_+ \end{array}\right)
 + \left(\begin{array}{cc} 0 & 0 \\ 0 & {\cal C}_+ \end{array}\right) \equiv
 \left(\begin{array}{cc} \hg^{uu}_+ & \hg^{ul}_+ \\ \hg^{lu}_+ & \hg^{ll}_+ \end{array}\right),
\end{equation}
where
\begin{equation} \label{Agammap}
 \left(\begin{array}{cc} \hga^{uu}_+ & \hga^{ul}_+ \\ \hga^{lu}_+ & \hga^{ll}_+ \end{array}\right) =
 \sum_{n>0,k_x,\epsilon}
 \frac{\psi_{nk_x\epsilon}\psi_{nk_x\epsilon}^{\dagger}}{\cZ-E_{nk_x\epsilon}}
\end{equation}
and
\begin{equation} \label{ACp0}
 \left(\begin{array}{cc} 0 & 0 \\ 0 & {\cal C}_+ \end{array}\right)
= \sum_{k_x,\epsilon} \frac{\psi_{0k_x\epsilon}\psi_{0k_x\epsilon}^{\dagger}}{\cZ-E_{0k_x\epsilon}}
 \delta_{\epsilon,-1}.
\end{equation}
For~$\hga^{uu}_+(\cZ)$ the summation over~$\epsilon=\pm 1$ gives
\begin{equation}
 \label{AGuu}
 \hga^{uu}_+(\cZ)=-\sum_{n>0,k_x}\left( \frac{\psi_+^u\psi_+^{u*}}{\cZ_+-E_n} +
 \frac{\psi_-^u\psi_-^{u*}}{\cZ_+ +E_n}\right),
\end{equation}
in which~$\cZ_+ = \cZ- E_{s_z}$, and~$\psi_{\pm}^{u}$ are defined
in Eqs.~(\ref{APsiP}) and~(\ref{APsiM}).
For~$\hga^{ul}_+$,~$\hga^{lu}_+$,~$\hga^{ll}_+$,~${\cal C}_+$
the summation over~$\epsilon=\pm 1$ is performed similarly.
Calculating in Eq.~(\ref{AGuu}) the integral over~$k_x$ we use
the identity, (see formula 7.377 in~\cite{GradshteinBook})
\begin{equation} \label{AeHmHn}
 \int_{-\infty}^{\infty}\!\!\!\! e^{-x^2}{\rm H}_m(x+y) {\rm H}_n(x+z)dx =
 2^n\sqrt{\pi}m!z^{n'}{\rm L}_m^{n'}(-2yz),
\end{equation}
where~$n'=n-m$,~$m\leq n$, and~$L_n^{\alpha}(t)$ are
the associated Laguerre polynomials. One gets
\begin{eqnarray}
\label{AGuu2}
 \hga^{uu}_+({\bm \rho},{\bm \rho}', \cZ) &=& -(\bcZ +\bEw)
 \sum_{n=1}^{\infty} \frac{A(\Omega)L_{n-1}^0(r^2)}{n + \bEw^2-\bcZ^2},\\
\label{AGll2}
 \hga^{ll}_+({\bm \rho},{\bm \rho}', \cZ) &=& -(\bcZ-\bEw)
 \sum_{n=1}^{\infty} \frac{A(\Omega) L_n^0(r^2)}{n + \bEw^2-\bcZ^2}, \\
 \label{ACp}
 {\cal C}_+({\bm \rho},{\bm \rho}', \cZ) &=&
 -(\bcZ-\bEw) \frac{A(\Omega) L_0^0(r^2)}{\bEw^2-\bcZ^2},\\
\label{AGul2}
 \hga^{ul}_+({\bm \rho},{\bm \rho}', \cZ) &=& m_{cd}
 \sum_{n=1}^{\infty} \frac{A(\Omega)L_{n-1}^1(r^2)}{n+ \bEw^2 - \bcZ^2},
\end{eqnarray}
where~$\bcZ=\cZ_+/(\hbar\Omega)$,~$\bEw=E_w/(\hbar\Omega)$,~$r^2 = ({\bm \rho} - {\bm \rho}')^2/(2L^2)$
and~$A(\Omega)$ is defined in Eq.~(\ref{g2D_A}).
In Eq.~(\ref{AGul2}) we defined~$m_{ul}= [(y-y')+i(x-x')]/(\sqrt{2}L)$.
For~$\hga^{lu}_+$ one obtains expression analogous to
that for~$\hga^{ul}_+$,
but with~$m_{ul}$ replaced by~$m_{lu} = [(y'-y)+i(x-x')]/(L\sqrt{2})$.

In Eqs.~(\ref{AGuu2})--(\ref{AGll2}) the summation over Landau levels~$n$ is performed
with the use of formulas~6.12.4 and~6.9.4 in~\cite{ErdelyiBook}
\begin{eqnarray}
 t^{-\beta} \sum_{n=0}^{\infty} \frac{L_n^{-\beta}(t)}{n+a-\beta}= \Gamma(a-\beta)\Psi(a,\beta+1;t) \nonumber \\
 \label{AFPsiW}
 =\Gamma(a-\beta) e^{t/2}t^{-1/2-\beta/2}W_{\beta/2+1/2-a,\beta/2}(t),
\end{eqnarray}
where~$\Psi(a,c;t)$ is the second solution of the confluent hypergeometric equation~\cite{ErdelyiBook},
and~$W_{\kappa,\mu}(t)$ is the Whittaker function.
The series in Eq.~(\ref{AFPsiW}) converges for~$t>0$ and~$\beta>-1/2$.
When combining Eqs.~(\ref{AGuu2})--(\ref{AFPsiW}) we set~$t=r^2$ and,
for~$\hga^{uu}_+$ and~$\hga^{ll}_+$ we take~$\beta=0$.
To calculate~$\hga^{uu}_+$ in Eq.~(\ref{AGuu2}) we first change the summation index~$n'= n-1$,
and then set~$a=1+\bEw^2- \bcZ^2$ in Eq.~(\ref{AFPsiW}).
Similarly, calculating the sum~$\hga^{ll}_+ + {\cal C}_+$ we set~$a=\bEw^2- \bcZ^2$.
Then we find
\begin{eqnarray}
 \label{AGuu3}
 \hg^{uu}_+ \equiv \hga^{uu}_+ &=& -\frac{(\bcZ+\bEw) e^{i\chi}}{2\pi\hbar\Omega L^2|r|}
 \cW_{\kappa_u}(r^2), \ \\
 \label{AGll3}
 \hg^{ll}_+ \equiv \hga^{ll}_+ + {\cal C}_+ &=& -\frac{(\bcZ-\bEw) e^{i\chi}}{2\pi\hbar\Omega L^2|r|}
 \cW_{\kappa_l}(r^2),
\end{eqnarray}
where~$\kappa_u= \bcZ^2 - \bEw^2 - 1/2$,~$\kappa_l = \bcZ^2 - \bEw^2 + 1/2$,
and~$\cW_{\kappa}(z) = \Gamma(1/2-\kappa) W_{\kappa,0}(z)$.
For~$\hga^{ul}_+$ there is~$\beta=-1$ in Eq.~(\ref{AFPsiW}), which is beyond the convergence
range of the series.
However,~$\hga^{ul}_+$ can be expressed as a
combination of~$\hga^{uu}_+$ and~$\hga^{ll}_+$ functions by using the
identity~$L_{n-1}^1(r^2)=n[(L_{n-1}^0(r^2)-L_{n}^0(r^2)]/r^2$,
see formula~8.971.4 in~\cite{GradshteinBook}. Then we obtain
\begin{equation} \label{AGul3}
\hg^{ul}_+ \equiv \hga^{ul}_+ =
 \frac{m_{ul}}{r^2}\left[(\bcZ-\bEw) \hg^{uu}_+ - (\bcZ+\bEw) \hg^{ll}_+ \right],
\end{equation}
where~$m_{ul}$ is defined in Eq.~(\ref{AGul2}).
Since GF is a hermitian operator,
there is~$\hg_+({\bm \rho},{\bm \rho}',E) = \langle {\bm \rho}|(E-\hH)^{-1}|{\bm \rho}'\rangle=
\hg_+({\bm \rho}',{\bm \rho},E)^{\dagger}$.
By interchanging~${\bm \rho}$ with~${\bm \rho}'$ in Eqs.~(\ref{AGuu3})--(\ref{AGul3}) one
finds:~$\hg^{cc}_+({\bm \rho},{\bm \rho}') = \hg^{cc}_+({\bm \rho}',{\bm \rho})^*$ with~$c\in\{u,l\}$,
and~$\hg^{ul}_+({\bm \rho},{\bm \rho}') = \hg^{lu}_+({\bm \rho}',{\bm \rho})^*$.
This verifies the hermiticity of GF in Eqs.~(\ref{AGuu3})--(\ref{AGul3}).

Now we show that, for~${\bm \rho} \rightarrow {\bm \rho}'$,
there is~$\hg^{ul}_+({\bm \rho},{\bm \rho}, \cZ) =\hg^{ul}_{{\bm 0},{\bm 0},+} = 0$.
To prove this we turn to Eq.~(\ref{AG22}) and after short algebra we find
\begin{equation} \label{AGulx1}
 \hg^{ul}_+({\bm \rho},{\bm \rho}', \cZ) = -\frac{\hbar\Omega}{2\pi}\sum_{n}\!\int_{-\infty}^{\infty}\!\!
 \frac{\sqrt{n} e^{ik_x(x-x')} \phi_m(\xi) \phi_n(\xi')^*}
 {\hbar^2 \Omega^2 n + E_w^2 -\cZ^2} dk_x,
\end{equation}
with~$m=n-1$. Taking the limit~${\bm \rho} \rightarrow {\bm \rho}'$ we have~$x=x'$ and~$\xi=\xi'$.
Then using~$d\xi = -Ldk_x$ we obtain
\begin{equation} \label{AGulx2}
 \hg^{ul}_+({\bm \rho},{\bm \rho}, \cZ) = \frac{\hbar\Omega}{2\pi L}\sum_{n}\!\int_{-\infty}^{\infty}\!\!
 \frac{\sqrt{n} \phi_m(\xi) \phi_n(\xi)}{\hbar^2 \Omega^2 n + E_w^2 -\cZ^2} d\xi = 0,
\end{equation}
because of the orthogonality of the Hermite functions~$\phi_m(\xi)$ and~$\phi_n(\xi)$ for~$n\neq m$.
For the same reason there is ~$\hg^{lu}_+({\bm \rho},{\bm \rho}, \cZ)=0$.
Since~$\hg^{ul}_+({\bm \rho},{\bm \rho}, \cZ)$ and~$\hg^{lu}_+({\bm \rho},{\bm \rho}, \cZ)$ vanish,
there is also~$\hg_{ul+}^{reg}=0$,~$\hg_{ul+}^{div}=0$,~$\hg_{lu}^{reg+}=0$ and~$\hg_{lu+}^{div}=0$,
see Eqs.~(\ref{gMLG_div})--(\ref{MLG_reg_ll}) and~(\ref{gVI_div_uu})--(\ref{gVI_reg_ll}).

For the~${\bm K'}$ point of BZ, the Hamiltonian~$\hH'$ is given in Eq.~(\ref{VI_hH2B}).
The stationary GF of the Hamiltonian~$\hH'$
is~$\left(\begin{array}{cc} \hg^{uu}_- & \hg^{ul}_-\\ \hg^{lu}_- & \hg^{ll}_- \end{array}\right)$,
see Eq.~(\ref{AG22}). Proceeding similarly as for the~${\bm K}$ point we find
\begin{eqnarray}
 \label{AGuu3p}
 \hg^{uu}_- &=& -\frac{(\bcZ'+\bEw) e^{i\chi}}{2\pi\hbar\Omega L^2|r|} \cW_{\kappa_u'}(r^2), \ \\
 \label{AGll3p}
 \hg^{ll}_- &=& -\frac{(\bcZ'-\bEw) e^{i\chi}}{2\pi\hbar\Omega L^2|r|} \cW_{\kappa_l'}(r^2), \\
 \label{AGul3p}
 \hg^{ul}_- &=& \frac{m_{ul}}{r^2}\left[(\bcZ'-\bEw) \hg^{ll}_- - (\bcZ'+\bEw) \hg^{uu}_- \right],
\end{eqnarray}
where~$\kappa_u'= \bcZ^{'2} - \bEw^{'2} + 1/2$,~$\kappa_l' = \bcZ^{'2} - \bEw^{'2} - 1/2$,
and~$\cW_{\kappa}(z) = \Gamma(1/2-\kappa) W_{\kappa,0}(z)$
with~$\cZ'=\cZ-E_{s_z}'$. The total GF is a~$8 \times 8$ block-diagonal matrix
consisting of four~$2\times 2$ blocks given in Eqs.~(\ref{AG22}) and~(\ref{AGul3p}) with
four combinations of~$\tau=\pm 1$ and~$s_z \pm 1$ quantum numbers.

\section{Numerical calculations of Whittaker functions}

One of the crucial elements in the calculations of Section~III is to use a fast
and accurate numerical procedure to obtain the Whittaker functions~$W_{\kappa,0}(z)$ and~$\cW_{\kappa}(z)$
appearing in the final formulas for the one-electron GF, see Eq.~(\ref{g2D_Ex}).
In this Appendix we collect numerical algorithms for
calculating~$W_{\kappa,0}(z)$ for various ranges of~$\kappa$ and~$z$.

It is useful to express GF in terms of the Whittaker functions~$W_{\kappa,0}(z)$,
as given in Eq.~(\ref{g2D_Ex}), because~$W_{\kappa,0}(z)$ can be conveniently
computed from the formula~9.237.1 in Ref.~\cite{GradshteinBook}:
\begin{eqnarray} \label{BW_Grad}
 W_{\kappa,0}(z) &=& \frac{\sqrt{z}\ e^{-z/2}}{\Gamma(1/2-\kappa)^2}
 \sum_{k=0}^{\infty}\frac{\Gamma(k-\kappa+1/2)}{(k!)^2}z^k \times \nonumber \\
 && \left[2\psi(k+1)-\psi(k-\kappa+1/2) -\ln(z)\right],
\end{eqnarray}
where~$\psi(z)=d\ln[\Gamma(z)]/dz$ and~$|{\rm arg}(z)| < 3\pi/2$.
In practical calculations it is helpful to calculate the
function:~$\cW_{\kappa}(z)= \Gamma(1/2-\kappa)W_{\kappa,0}(z)$.
In this case, applying~$k$ times
the identity:~$\Gamma(1/2-\kappa+k) =(1/2-\kappa+k-1)\Gamma(1/2-\kappa+k-1)$,
one avoids the direct calculation of~$\Gamma(1/2-\kappa)$ function,
which simplifies calculations of~$\cW_{\kappa}(z)$ for large~$\kappa$.
This method gives stable numerical results
within the rectangular:~$ -5 < {\rm Re} (\kappa) < 25$ and~$0<z<25$.
For~${\rm Re} (\kappa)$ outside the above region,
the formula in Eq.~(\ref{BW_Grad}) gives erroneous results due to
truncation errors. To avoid these problems,
we devised a numerically-stable iterative algorithm of calculating
the Whittaker function for large values of~${\rm Re} (\kappa)$.

Consider the identity~13.4.31 in Ref.~\cite{AbramowitzBook} for~$W_{\kappa,0}(z)$
\begin{equation} \label{BW_1}
 W_{\kappa+1,0}(z) = - (2\kappa-z)W_{\kappa,0}(z) -(\kappa-1/2)^2 W_{\kappa-1,0}(z).
\end{equation}
On multiplying both sides
by~$\Gamma(1/2-\kappa-1)(-1)^{{\rm Re}(\kappa)+1}$ and
using three times the identity~$\Gamma(x+1)=x\Gamma(x)$ one obtains
\begin{equation} \label{BW_Stable}
 \cW_{\kappa+1}^{-}(z) = - \frac{2\kappa-z}{\kappa+1/2} \cW_{\kappa}^{-}(z)
 -\frac{\kappa -1/2}{\kappa+1/2} \cW_{\kappa-1}^{-}(z),
\end{equation}
where~$\cW_{\kappa}^{-}(z) = \cW_{\kappa}(z)(-1)^{{\rm Re}(\kappa)}$.
Equation~(\ref{BW_Stable}) allows one to calculate
iteratively~$\cW_{\kappa}^{-}(z)$ for~$z>0$ and large~${\rm Re}(\kappa) >0$ values.
To show the stability of the iterative scheme in Eq.~(\ref{BW_Stable}) we consider
the limit of both sides of Eq.~(\ref{BW_Stable}) for large~$\kappa$ values
\begin{equation} \label{BW_Large}
 \cW_{\kappa+1}^{-}(z) \simeq \left(2- \frac{z}{\kappa}\right)\cW_{\kappa}^{-}(z)
 -\cW_{\kappa-1}^{-}(z).
\end{equation}

Consider now a function~$f(\kappa)$ satisfying the second-order differential
equation:~$f^{"}(\kappa) = -\omega^2f(\kappa)$.
Then~$f(\kappa)= C\sin(\omega \kappa +\phi)$, where the constants~$C$ and~$\phi$
are determined from the initial conditions.
Approximating the second derivative~$f^{"}(\kappa)$ by a finite
difference one obtains
\begin{equation} \label{B_Sin}
 f_{\kappa+1} = (2 - \omega^2h^2) f_{\kappa} - f_{\kappa-1}.
\end{equation}
The iteration scheme in Eq.~(\ref{B_Sin}) is numerically stable
for~$(\omega^2 h^2) < 1$, which may be verified by direct calculations.

By comparing Eq.~(\ref{B_Sin}) with Eq.~(\ref{BW_Large}),
taking~$h= 1$ and~$\omega=\sqrt{z/\kappa}$ we find that i)
for large~${\rm Re}(\kappa)$ the iteration scheme in Eq.~(\ref{BW_Large})
is numerically stable for~$(z/\kappa) < 1$, and
ii) for large~$\kappa$ values there
is:~$\cW_{\kappa+1}^{-}(z) \simeq A\sin(\sqrt{z\kappa} + \Phi)$,
i.e.~$\cW_{\kappa}^{-}(z)$ oscillates with a finite amplitude.

The factor~$(-1)^{{\rm Re}(\kappa)}$ in definition
of~$\cW_{\kappa}^{-}(z)$ plays the key role in the
iteration scheme in Eq.~(\ref{BW_Large}), since it ensures the negative sign
in front of~$(z/\kappa)$ term.
If one iterated~$\cW_{\kappa}(z)$ instead of~$\cW_{\kappa}^{-}(z)$
and the factor~$(-1)^{{\rm Re}(\kappa)}$ were omitted,
the corresponding sign in front of~$(z/\kappa)$ term in Eq.~(\ref{BW_Large})
would be positive and the iterating scheme would diverge.

The iterative procedure in Eq.~(\ref{BW_Large}) works correctly for~$\kappa > 0$.
For small negative values of~$\kappa$ or~${\rm Re}(\kappa)$ one can still
use Eq.~(\ref{BW_Grad}), while for large negative~$\kappa$ one can
use formula~9.229.2 in Ref.~\cite{GradshteinBook}
\begin{equation} \label{BW_neg}
 W_{-|\kappa|,0}(z) \simeq \left(\frac{z}{4|\kappa|}\right)^{1/4}
 e^{|\kappa| - |\kappa| \ln(|\kappa|)}e^{-2\sqrt{|\kappa| z}}.
\end{equation}
The above formula describes the exponential decay of the Whittaker functions for
large complex~$z$ or large negative~$\kappa$.

\section{Density of states from Green's function~$\hg^{reg}$}

In our model we use Lorentz profile of the energy levels in a magnetic field.
We claimed that such a profile follows directly from the
regularization procedure. Here we prove this statement.
Turning to Eq.~(\ref{g2D_reg}) we assume that~$\bcE$ is a complex
number:~$\bcE= \bcE + i\eta$, where~$\eta$ is small but finite.
By substituting~$\bcE$ to~$\hg^{reg}$ in Eq.~(\ref{g2D_reg}) and
using~$n$ times the formula:~$\psi(z+1) = \psi(z)+1/z$, we find
that in the vicinities of~$\bcE \simeq n+1/2$ there is
(see formula~1.17.12 in Ref.~\cite{ErdelyiBook} with~$m=0$ or
Ref.~\cite{WoframPolyGamma})
\begin{equation} \label{g2D_reg_a}
 \hg^{reg}= \simeq c_0\psi(i\eta) + \frac{c_0}{(\bcE - n-1/2) + i\eta},
\end{equation}
where~$c_0$ is a constant, see Eq.~(\ref{g2D_reg}).
For nonzero~$\eta$ the first term in Eq.~(\ref{g2D_reg_a}) is
finite:~$\psi(i\eta) \simeq -1/(i\eta) -\gamma + \ldots$, so that
for~$\hg^{reg}$ in Eq.~(\ref{g2D_reg_a})
the DOS obtained from~${\rm Im}\{\hg^{reg}\}$ for
energies~$\bcE$ close to~$n+1/2$ is a Lorentz function
centered at~$\bcE= n+1/2$.

The Lorentz-like shape of Landau levels DOS is less commonly encountered
than the Gaussian one, but it was used for monolayer graphene
in Refs.~\cite{Sharapov2004,Jiang2007}.
The calculations in Section~III depend weakly on the shape of Landau level DOS and
its width~$\eta$ as long as the consecutive levels do not overlap with each other.
This condition is met for 2D electron gas for sufficiently small~$\eta$,
but it fails for large LL numbers~$n$ in the two remaining systems.
However for material parameters used in our paper this problem
appears for~$n \simeq 12$, which exceeds Fermi energies used in Section~III.

\section{Finite width of Green's function}

\begin{table}
\label{Table_2}
\caption{Hilbert transforms~$H(y)=(1/\pi) {\cal P} \int_{-\infty}^{\infty} dy( f(x)/ (x-y))$ of bell-like
        functions~$f(x)$ used in modeling DOS for electron gas in the presence of scattering.
        Notation:~$y_n=E-E_n$ is used.}
\begin{tabular}{|c|c|c|}
 \hline
 function &~${\rm Im} \{\hg_{\bm 0,\bm 0}\}$ &~${\rm Re}\{\hg_{\bm 0,\bm 0}\}$ \\
 \hline
 Gaussian & \(\displaystyle \frac{1}{\sqrt{\pi}} \sum_n e^{-(E-E_n)^2} \) & \(\displaystyle \frac{2}{\pi}\sum_n D_+(y_n) \) \\
 \hline
 Lorentz & \(\displaystyle \frac{1}{\pi} \sum_n \frac{1}{(E-E_n)^2+1} \) & \(\displaystyle \frac{1}{\pi} \sum_n \frac{y_n}{y_n^2+1} \) \\
 \hline
 Rectangular & \(\displaystyle \sum_n {\rm rect} (E-E_n) \) & \(\displaystyle \frac{1}{\pi} \sum_n \ln \left| \frac{y_n+1/2}{y_n-1/2} \right| \) \\
 \hline
 Sinc & \(\displaystyle \frac{1}{\pi} \sum_n \frac{\sin(E-E_n)}{E-E_n} \) & \(\displaystyle \frac{1}{\pi} \sum_n\frac{1 - \cos(y_n) }{y_n} \) \\
 \hline
 Semi-circle & \(\displaystyle \frac{1}{\pi} \sum_n \sqrt{a^2-(E-E_n)^2} \) & \(\displaystyle \sum_n \frac{y_n}{|y_n|}\sqrt{y_n^2-a^2} - y_n \) \\
             &                                                              & \(\displaystyle \sum_n - y_n, \ |y_n| < a \) \\
 \hline
\end{tabular}
\end{table}

In this Appendix we analyze a possible impact of several shapes of Landau levels
on divergencies of~$\hg_{\bm 0,\bm 0}$ in Eq.~(\ref{g2D_r0}).
We concentrate on 2D electron gas. Calculations for electrons in monolayer graphene
and group-VI dichalogenides are similar to those presented below.
It is suggested that for {\it any} bell-like DOS the
divergencies~$\hg_{\bm 0,\bm 0}$ occur because
one finally obtains a divergent harmonic series for~$\hg_{\bm 0,\bm 0}$.
This statement is not proven in general, but the examples supporting it
are listed and discussed below.

Let us consider the Green's function~$\hg = 1/(E-\hH_0 - i\eta)$,
where~$\eta$ is small but finite.
In the spatial representation~$\hg$ is given in Eq.~(\ref{g2D_Ex}).
For sufficiently small~$\eta$ we may write
\begin{equation} \label{C_g00a}
 \hg_{\bm 0,\bm 0} = \left\langle{\bm \rho}_0\right|{\cal P}\left(\frac{1}{E-\hH_0}\right)
 + i\pi \delta\left(E-\hH_0\right)\left|{\bm \rho}_0\right\rangle.
\end{equation}
On inserting the complete set of eigenstates~$|{\rm n}\rangle$ of~$\hH_0$ in
the RHS of~(\ref{C_g00a}) we obtain
\begin{equation} \label{C_g00b}
 \hg_{\bm 0,\bm 0} = \sum_{\rm n} \Psi_{\rm n}^{\dagger} \Psi_{\rm n}
 {\cal P} \left(\frac{1}{E-E_{\rm n}}\right)
 + i\pi \sum_{\rm n} \Psi_{\rm n}^{\dagger} \Psi_{\rm n} \delta(E-E_{\rm n}),
\end{equation}
in which~$\hH_0|{\rm n}\rangle = E_{\rm n}|{\rm n}\rangle$.
Here~${\rm n}$ denotes all quantum numbers
describing the eigenstate~$|{\rm n}\rangle$,
while~$\Psi_{\rm n} \equiv \Psi_{\rm n}({\bm \rho}_0) = \langle {\bm \rho}_0|{\rm n}\rangle$
is the eigenfunction of~$\hH_0$ in the position representation. For 2D electron
gas there is:~${\rm n}=|nk_x\rangle$,~$E_n=\hbar\omega_c(n+1/2)$,
and~$\Psi_{\rm n}$ are given in Eq.~(\ref{2D_Psi}).

The DOS of the system is proportional to~${\rm Im} (\hg_{\bm 0,\bm 0})$, see Eq.~(\ref{n_rho}).
In absence of scattering~${\rm DOS} \propto \sum_{\rm n}\delta(E-E_n)$.
In the presence of scattering the DOS peaks have finite widths~$\eta$ and
finite height. There exist several models of DOS in the literature and
the most common is the Gaussian form
\begin{equation} \label{CDos}
 {\rm DOS} =\frac{1}{\sqrt{\pi}\ \eta} \sum_{\rm n}e^{-(E-E_n)^2/\eta^2}.
\end{equation}
For this choice we may {\it assume} that
\begin{eqnarray} \label{CImg1}
 {\rm Im} \hg_{\bm 0,\bm 0} &=&\pi \sum_{\rm n} \Psi_{\rm n}^{\dagger} \Psi_{\rm n} \delta(E-E_n) \nonumber \\
 &\rightarrow & \frac{\pi}{\sqrt{\pi}\ \eta} \sum_{\rm n}\Psi_{\rm n}^{\dagger} \Psi_{\rm n}e^{-(E-E_n)^2/\eta^2}.
\end{eqnarray}
Let us analyze consequences of such assumption. Since the eigenenergies~$E_n$ do not depend on~$k_x$,
the sum in Eq.~(\ref{CImg1}) is
\begin{equation} \label{CImg2}
 {\rm Im} \hg_{\bm 0,\bm 0}= \frac{\pi}{\sqrt{\pi}\eta}\sum_n \hat{Q}_n e^{-(E-E_n)^2/\eta^2},
\end{equation}
where
\begin{equation} \label{CQ}
 \hat{Q}_n =\int_{-\infty}^{\infty} \Psi_{nk_x}^{\dagger}(\bm \rho_0) \Psi_{nk_x}(\bm \rho_0) dk_x.
\end{equation}
Using Eq.~(\ref{2D_Psi}) and calculating
the integral over~$k_x$ one finds~$\hat{Q}_n=1/(2\pi L^2)$ for all~$n$.

The real part of~$\hg_{\bm 0,\bm 0}$ is related to the
imaginary part by the Hilbert transform
\begin{equation} \label{CHilbert0}
 {\rm Re} \{\hg_{\bm 0,\bm 0}\} = \frac{1}{\pi}{\cal P} \int_{-\infty}^{\infty}
 \frac{{\rm Im} \{\hg_{\bm 0,\bm 0}(x)\}}{x-E} dx.
\end{equation}
Thus the replacement of the Dirac deltas for the imaginary part of~$\hg_{\bm 0,\bm 0}$
by the Gaussian function requirers an appropriate modification of the real part
\begin{equation} \label{CReg1}
{\rm Re} \{\hg_{\bm 0,\bm 0}\} = \left(\frac{1}{2\pi L^2}\right)
{\cal P} \int_{-\infty}^{\infty} \sum_n \frac{e^{-(x-E_n)^2/\eta^2}}{\sqrt{\pi}\eta(x-E)}dx.
\end{equation}
We assume for a moment that the summation over~$n$ is truncated to a finite~$N_{max}$,
so one can change the order of summation and the integration.
Then one finds that the integral in Eq.~(\ref{CReg1})
describes the Hilbert transform of the Gaussian function~$u(t)=e^{-t^2}$,
which is:~$H(s)= 2/\sqrt{\pi}D_+(s)$, where~$D_+(s)$ is the Dawson function
(see formulas~7.1.3 and 7.1.4 in Ref.~{\cite{AbramowitzBook}}). This gives
\begin{equation} \label{CReg2}
 {\rm Re} \{\hg_{\bm 0,\bm 0}\} = \frac{1}{\pi^2 L^2 \eta} \sum_n^{N_{max}} D_+(E/\eta-E_n/\eta).
\end{equation}
For large arguments, the Dawson function decays as~$(E-E_n)^{-1}$,
which leads to a divergence of the
sum in Eq.~(\ref{CReg2}) for~$N_{max} \rightarrow \infty$,
since in this limit one obtains the harmonic series.

In Table~2 we listed five bell-like functions used in the literature for calculations of
DOS in 2D systems. The rectangular function is defined
as:~${\rm rect}(x)=1$ for~$|x|< 0.5$,~${\rm rect}(x)=0$ for~$|x|>0.5$,
and~${\rm rect}(x)=0.5$ for~$|x|=0.5$. We set~$\eta=1$ and denote~$y_n=E-E_n$.
Real parts~${\rm Re} \{\hg_{\bm 0,\bm 0}\}$ are obtained by the Hilbert transforms
of~${\rm Im} \{\hg_{\bm 0,\bm 0}\}$, see Eq.~(\ref{CHilbert0}).
In all cases, for large~$n$ the functions~${\rm Re}\{\hg_{\bm 0,\bm 0}\}$
reduce to the harmonic series~$\sum_n 1/y_n = \sum_n (E-E_n)^{-1}$
and diverge as~$(E-E_n)^{-1}$.
The conclusion from Table~2 is that, for electrons in a magnetic field,
{\it any} reasonable bell-like form of DOS encountered in the literature
leads to divergences of~${\rm Re}\{\hg_{\bm 0,\bm 0}\}$.
This seems to be an unavoidable feature of the problem.
The divergence should be eliminated using other methods as, e.g.
the regularization procedure described in Section~II.

\hspace*{1em}

\end{document}